\documentclass[12pt]{iopart}
\usepackage{ifpdf}
\ifpdf
   \usepackage[pdftex]{graphicx}
   \DeclareGraphicsExtensions{.pdf}
   \usepackage{thumbpdf}      %%% thumbnails for pdflatex
\else
   \usepackage[dvips]{graphicx}
   \DeclareGraphicsExtensions{.eps}
\fi
\usepackage{cite}
\usepackage{amssymb}
\usepackage{amscd}
\usepackage{subfigure}
\usepackage{setstack}
\usepackage{iopams}

\renewcommand{\text}[1]{{\mbox{#1}}}
\renewenvironment{pmatrix}{\left(\!\!\begin{array}{cc}}{\end{array}\!\!\right)}
\newcommand{\ri}{{ \rm i }}

\newcommand{\rd}{{ \rm d }}
\newcommand{\rR}{{ \rm R }}
\newcommand{\rC}{{ \rm C }}

\newcommand{\rt}{{ \rm t }}
\newcommand{\sk}{{ \rm sk }}
\newcommand{\tx}{{ \tilde x }}
\newcommand{\tu}{{ \tilde u }}

\newcommand{\be}{\begin{equation}}
\newcommand{\ee}{\end{equation}}

\newcommand{\sgn}{{\rm sgn}}
\newcommand{\cn}{{\rm cn}}
\newcommand{\sn}{{\rm sn}}
\newcommand{\dn}{{\rm dn}}

\usepackage{color}
\definecolor{blau}{rgb}{0,0,1}
\definecolor{gruen}{rgb}{0,1,0}
\definecolor{rot}{rgb}{1,0,0}
\definecolor{magenta}{rgb}{1,0,1}
\begin{document}
\jl{1}
\title[Multi-barrier resonant tunneling for the one--dimensional NLSE]
{Multi-barrier resonant tunneling for the one-- dimensional nonlinear Schr\"odinger Equation}
\author{K Rapedius and H J Korsch}

\address{ FB Physik, Universit\"at Kaiserslautern, D-67653
Kaiserslautern, Germany}

\ead{korsch@physik.uni-kl.de}

\begin{abstract}
For the stationary one-dimensional nonlinear Schr\"odinger equation (or Gross-Pitaevskii equation)
nonlinear resonant transmission through a finite number of equidistant identical barriers is studied using a (semi--) analytical approach.
In addition to the occurrence of bistable transmission peaks known from nonlinear resonant transmission through a single quantum well (respectively a double barrier) complicated (looped) structures are observed in the transmission coefficient which can be identified as the result of symmetry breaking similar to the emergence of self-trapping states in double well potentials. 
Furthermore it is shown that these results are well reproduced by a nonlinear oscillator model based on a small number of resonance eigenfunctions of the corresponding linear system.
\end{abstract}

%\submitto{\JPA}
\label{chap-Multi}
%In the chapters ... we considered nonlinear resonant tunnelling through single well/double barrier structures finding effects like bistability of transmission in the vicinity of a resonance. 

\section{Introduction}
Transport properties of Bose-Einstein condensates (BECs) in (quasi-) one-dimensional waveguides are of
considerable current interest, both experimentally and theoretically. Especially atom--chip experiments %\cite{Folm00,Hans01,Ott01,Ande02}
are well--suited to study the influence
of interatomic interaction on transport properties of BECs in waveguides since various waveguide geometries can be realized by different methods \cite{Folm00,Hans01,Ott01,Ande02,Lean02,Vale04,Sinc05,Brav06,Guer06,Sing08}.
%using different techniques, as for instance magnetic  \cite{Lean02,Sinc05,Sing08}, photonic crystal \cite{Brav06} or foil based chips \cite{Vale04}.
%An alternative method was implemented in a recent experiment \cite{Guer06} where a BEC was created in an optomagnetic trap and outcoupled into an optical waveguide.

A convenient theoretical approach is based on the one-dimensional Gross-Pitaevskii equation (GPE)
or nonlinear Schr\"odinger equation (NLSE)
\be
   \ri \hbar \dot{\psi}(x,t)=\left( -\frac{\hbar^2}{2m} \frac{\rd^2}{\rd x^2}+g|\psi(x,t)|^2+V(x) \right) \psi(x,t)
   \label{GPE}
\ee
which describes the dynamics in a mean-field approximation at low
temperatures \cite{Pita03,Park98,Dalf99,Legg01}. The nonlinear term $g|\psi(x,t)|^2$ models the interaction between the condesate particles. Another important application of the NLSE is the propagation of electromagnetic waves in
nonlinear media (see, e.g., \cite[Ch.8]{Dodd82}).
The ansatz $\psi( x,t)=\exp(-\ri \mu t/\hbar)\,\psi(x)$ reduces
(\ref{GPE}) to the
corresponding time-independent NLSE
\be
   \left( -\frac{\hbar^2}{2m} \frac{\rd^2}{\rd x^2}+g|\psi(x)|^2+V(x)-\mu \right) \psi(x)=0
   \label{GPE_1D_stat}
\ee
with the chemical potential $\mu$.

Within this framework resonant transport through single well (respectively double barrier) structures has been studied using analytical and numerical approaches \cite{Paul05,Paul05b,06nl_transport}. It was found that due to the nonlinearity of equation (\ref{GPE_1D_stat}) the barrier transmission coefficient in depencence on the chemical potential $\mu$ shows bistable resonance peaks which can be related to nonlinear metastable (Siegert) resonance states of the barrier potential and described by means of a nonlinear generalization of the  Lorentzian profile occuring in linear transmission problems \cite{Paul05b,08nlLorentz}.
These studies correspond to the barrier tunneling of coherent monochromatic matter waves with a given chemical potential $\mu$ that are injected into the waveguide from a BEC reservoir. In the articles cited above it was shown that the results obtained from the stationary NLSE (\ref{GPE_1D_stat}) are in excellent agreement with numerical solutions of the time--dependent NLSE
\be
\fl
   \quad \ri \hbar \dot{\psi}(x,t)=-\frac{\hbar^2}{2m}\psi''(x,t)+V(x)\psi(x,t)
   +g|\psi(x,t)|^2\psi(x,t)+f_0\exp(-\ri \mu t/\hbar)\delta(x-x_0)\,
   \label{GPE_t}
\ee
where the coupling to a reservoir is modeled by the source term $f_0 \exp(-\ri \mu t/\hbar)\delta(x-x_0)$ located at some position $x=x_0$ on the left hand side of the barrier (i.~e.~in the upstream region), emitting monochromatic matter waves at chemical potential $\mu$. In contrast to the linear case these results cannot be straightforwardly used to predict the scattering behaviour of an arbitrary wavepacket since the superposition principle is no longer valid.% and thus simulates the coupling to a reservoir.

On the other hand double well potentials have been considered in a number of theoretical and experimental papers (see e.~g.~\cite{Albi05,Schu05b,Infe06,Gati06,Gati06b,Theo06,Khom07,Foel07,09ddshell}). In such systems one observes the onset of symmetry breaking and the emergence of new solutions in addition to the solutions with linear counterpart for a critical value of the nonlinearity. These results strongly motivate a study of related effects in the context of resonant transmission through structures consisting of more than one well (or more than two barriers, respectively) where one expects the occurrence of both bistability and symmetry breaking. 
The limiting case of resonant transport in infinitely extended periodic structures has also  been of recent interest (see, e.~g.~\cite{Seam05b,Wu03,09dcomb}). The occurrence of looped Bloch bands is one of the major effects of nonlinearity in these systems. Transport through a finite number of delta barriers was considered in \cite{Tara99}, however, focusing on different aspects like the superfluidity of the condensate flow. In this paper we thus intend to fill a gap by considering nonlinear resonant tunnelling through a finite sequence of $n$ identical equidistant barriers. For the linear Schr\"odinger equation transmission through such a multi-barrier or truncated periodic potential has been investigated in a number of theoretical papers motivated both by experiments with multilayered semiconductor heterostructures as well as fundamental issues like their relationship to infinitely extended periodic potentials. The particular systems treated in the literature include analytically solvable potentials like sequences of rectangular \cite{Yama90,Yong91} or delta-function barriers \cite{Grif92,Kian74,Yong91}.

In the following we consider resonant transmission for the stationary one-dimensional NLSE (\ref{GPE_1D_stat})
with the potential
\be
   V(x)=\frac{\hbar^2}{m}\lambda \sum_{j=0}^{n-1} \delta(x-jd) \label{delta_series}
\ee
consisting of $n$ identical delta-function barriers with distance $d$ and strength $\hbar^2 \lambda/m$ with $\lambda>0$.

This paper is organized as follows. In section \ref{sec_Multi_intro} we have a brief look at the potential (\ref{delta_series}) the linear limit $(g=0)$. In section \ref{sec-Multi_transfer} we introduce a semi--analytical method for calculating the transmission coefficient for piecewise constant potentials which is applied to the case of double, triple, quadruple and quintuple barrier tunnelling in section \ref{sec-Multi_Tq}. 
In section \ref{sec_Multi_NLO} these results are compared with the predictions of a nonlinear oscillator model.
Additional material concerning computational details is presented in an appendix.
%--------------------------------------------------------------------------------------------------------------------------------------
\section{Multi barrier transmission in the linear limit}
\label{sec_Multi_intro}
\begin{figure}[htb]
\centering
%\psfrag{m}[c][c]{{\footnotesize ${\displaystyle {\cal E}_{\rm Test}}$}}
\includegraphics[width=7cm,  angle=0]{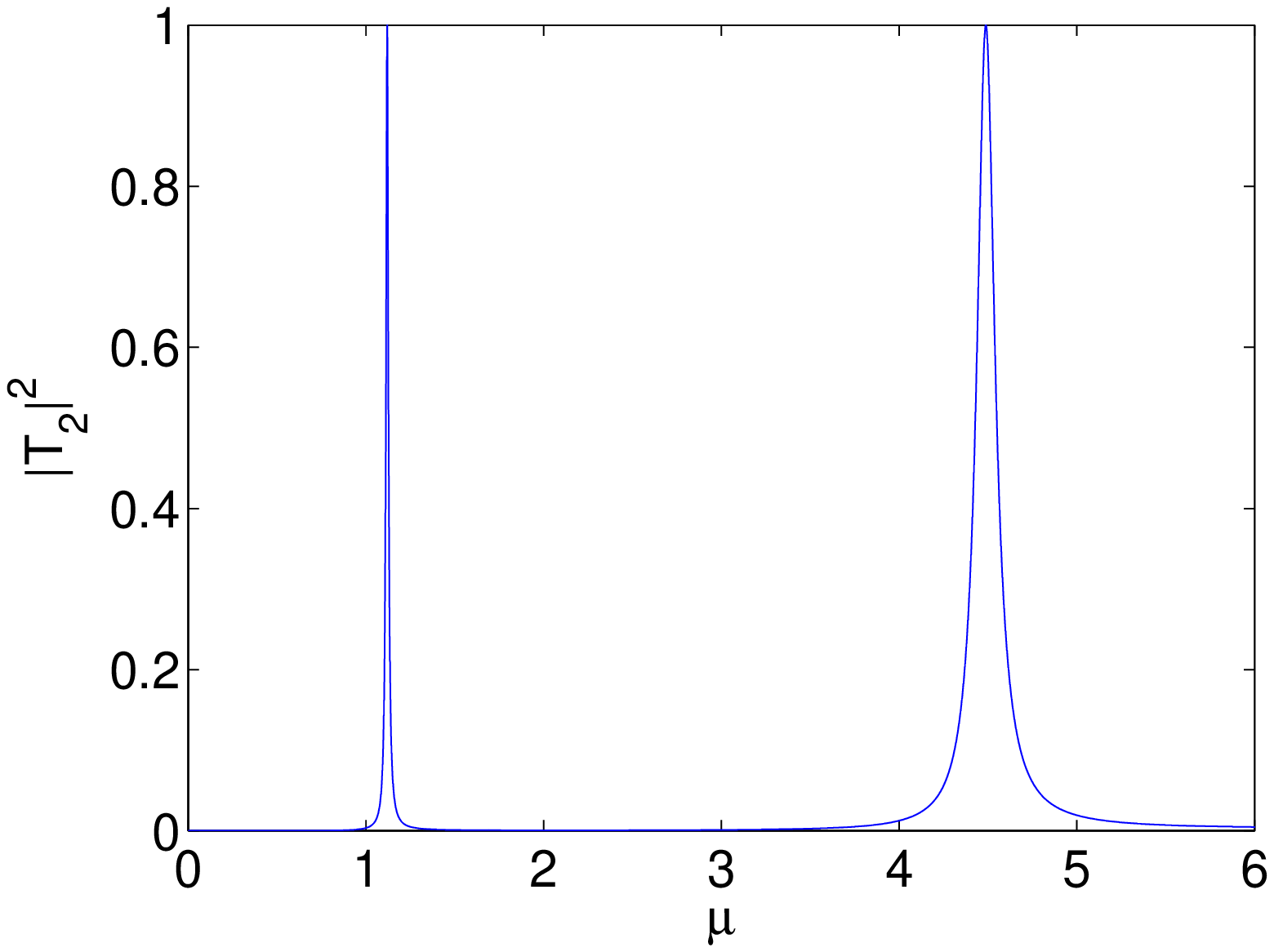}
\includegraphics[width=7cm,  angle=0]{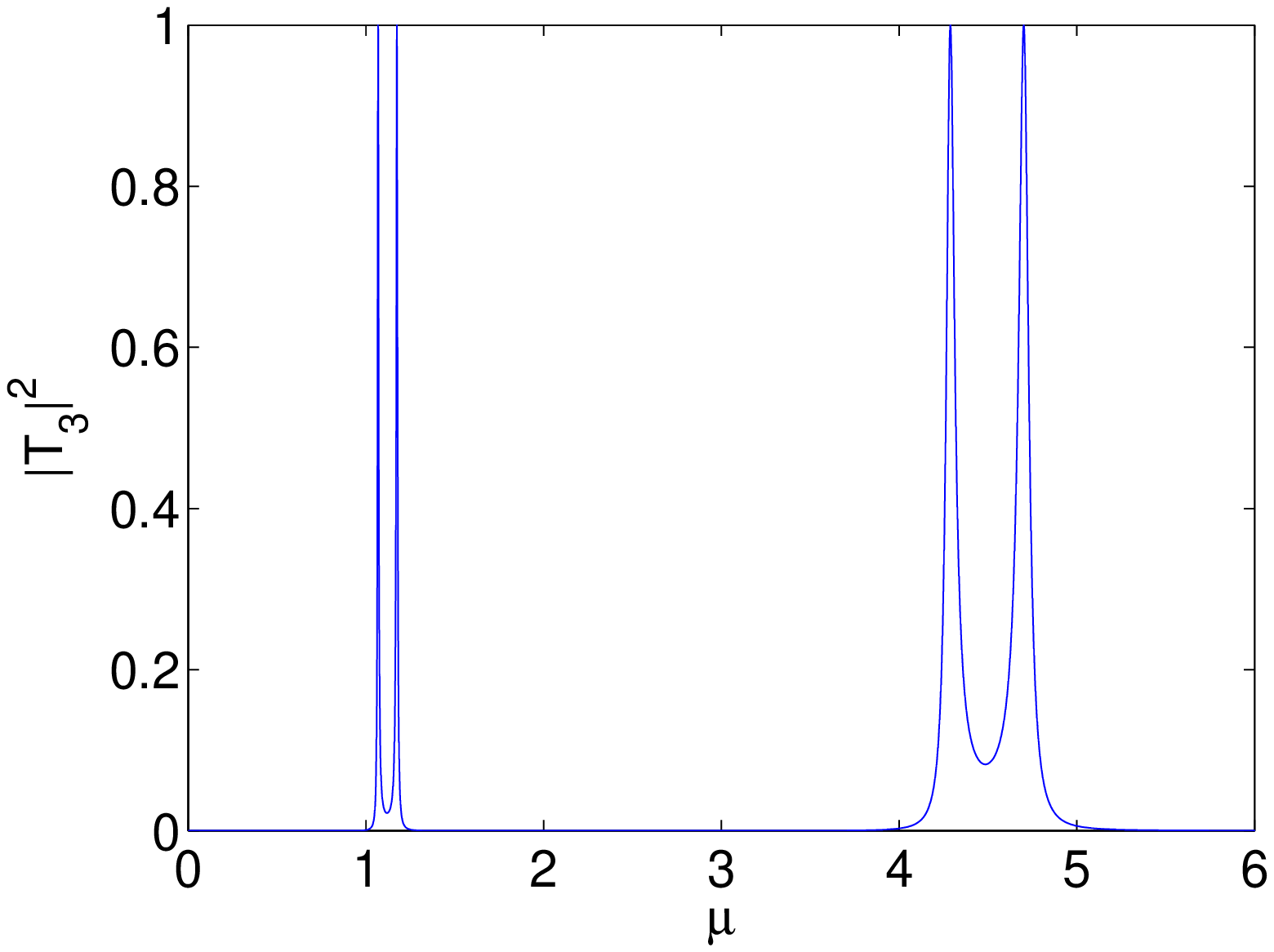}\\
\includegraphics[width=7cm,  angle=0]{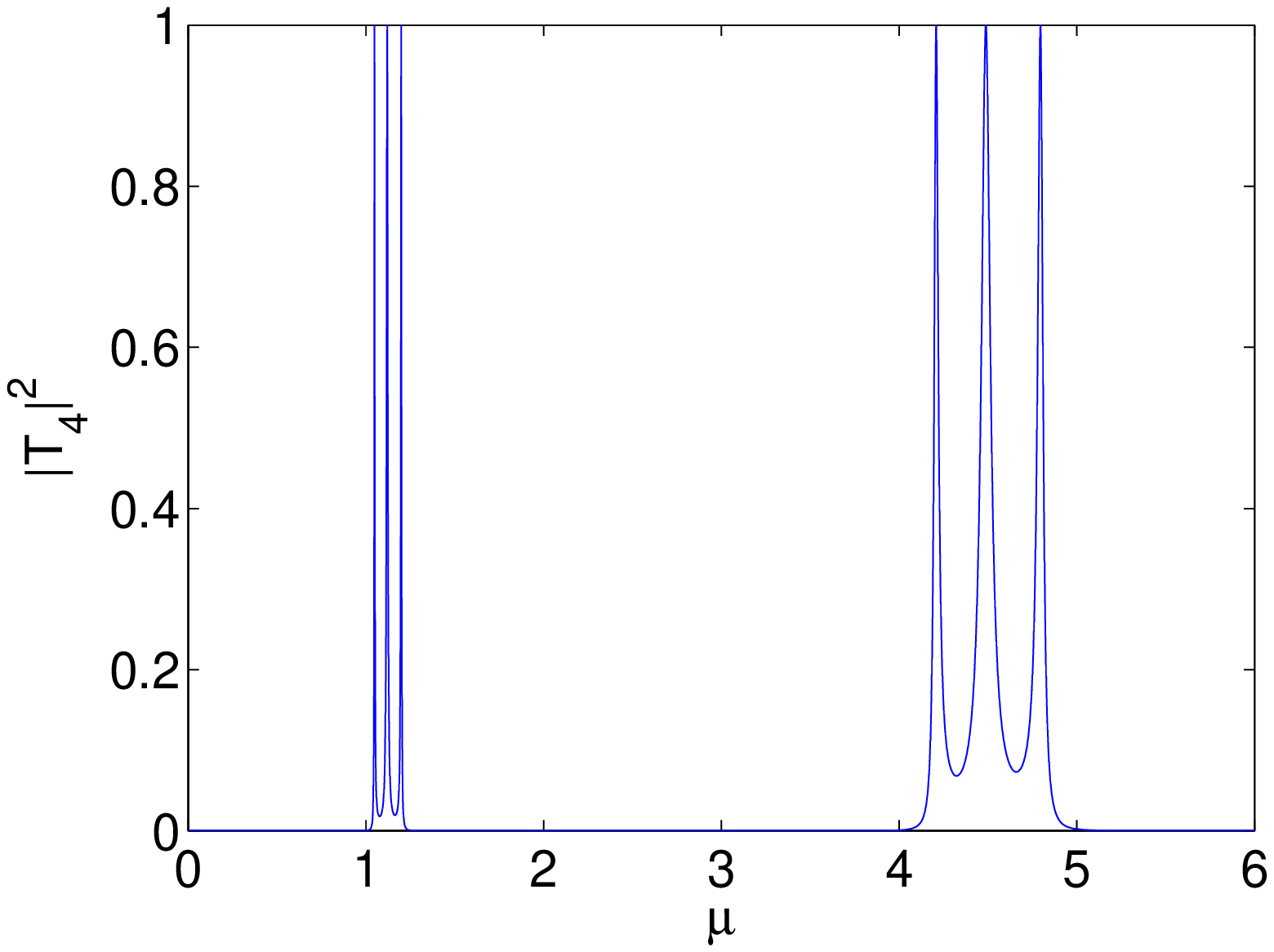}
\includegraphics[width=7cm,  angle=0]{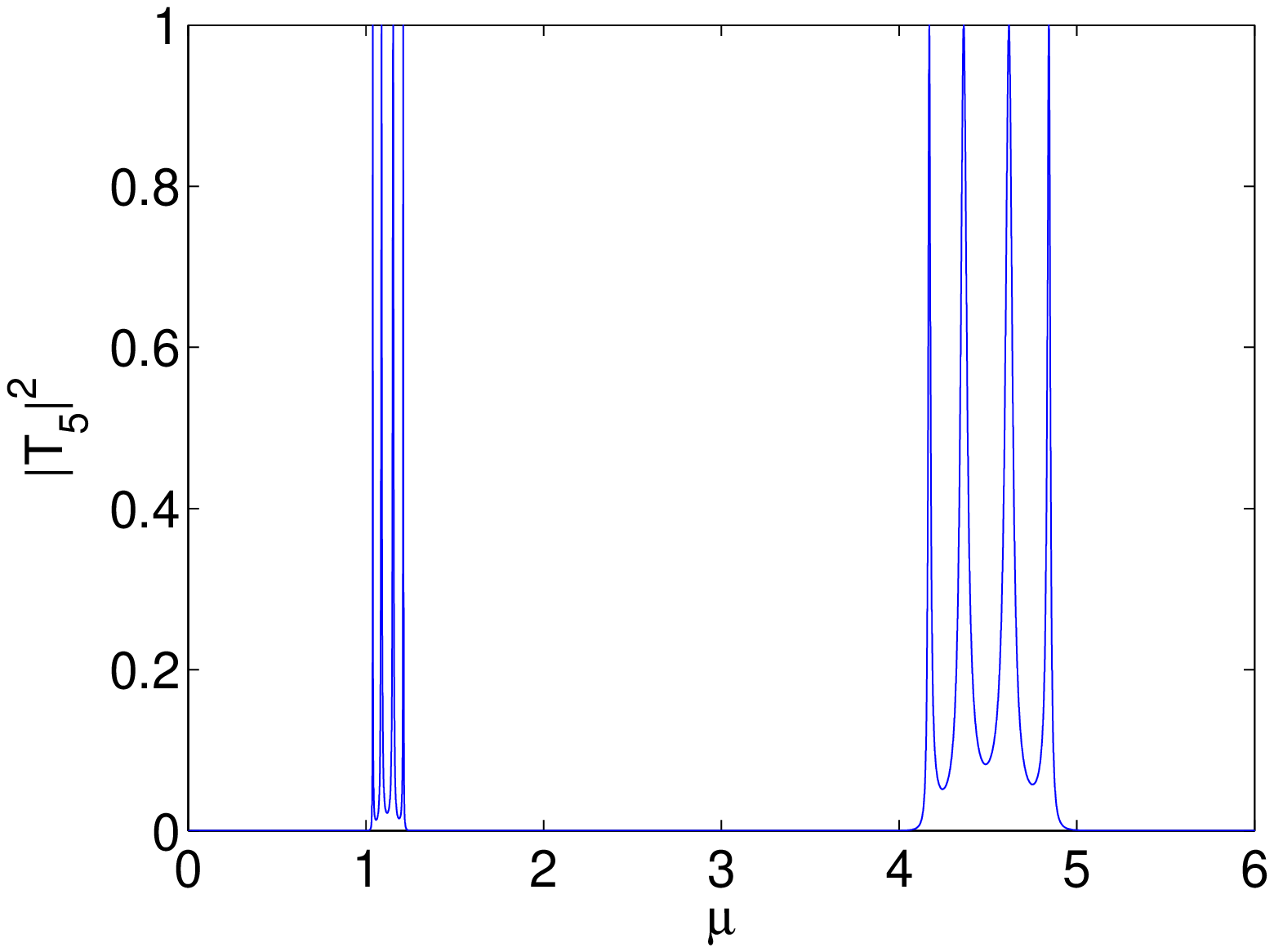}\\
\caption{\label{fig-linT2_5} {Transmission coefficients $|T_n|^2$ in dependence on the chemical potential $\mu$ for $n=2,3,4,5$ and the parameters $\lambda=10$, $d=2$, $g=0$.}}
\end{figure}

In this section we briefly discuss transmission through the barrier potential (\ref{delta_series}) for the linear Schr\"odinger equation, i.~e.~equation (\ref{GPE_1D_stat}) with $g=0$.
Using the transfer matrix technique, Griffiths and Taussig \cite{Grif92} have proven that the transmission coefficient for $n$ delta barriers is given by
\be
  |T_n(\mu)|^2=[1+(\lambda/k)^2 U^2_{n-1}(z)]^{-1}
   \label{Tq_lin}
\ee
with $k=\sqrt{2m\mu}/\hbar$,
\be
  z= \cos kd + (\lambda/k)\sin kd \,
  \label{z}
\ee
and the Chebyshev polynomials $U_n(z)$ of the second kind generated by the recurrence relation
\be
   U_{n+1}(z)=2z U_{n}(z)-U_{n-1}(z)
\ee
starting with $U_0(z)=1$ and $U_1(z)=2z$. Note that for any given value of $z$ with $-1 < z < 1$ equation (\ref{z}) yields infinitely many solutions $k$ (respectively $\mu$). The resonant chemical potentials $\mu_\rR$ where $|T_n(\mu_\rR)|^2=1$ can be determined by solving the transcendental equation (\ref{z}) with the roots
\be
   z_l=\cos(l \pi/n),\quad l=1,\dots,n-1
   \label{z_l}
\ee
of the Chebyshev polynomial $U_{n-1}(z)$. Since $U_{n}(z)$ has $n$ zeros, $|T_1|^2$ has no resonances and the resonances of $|T_n|^2$ with $n \ge 2$ occur in groups of multiplicity $n-1$. Because of $l/n = (\nu l)/(\nu n)$ with $\nu=1,2, \dots$ we see from equation (\ref{z_l}) that any resonance of $|T_n|^2$ is also a resonance of $|T_{\nu \cdot n}|^2$.
This result can be understood in an intuitive way by decomposing a series of $\nu \cdot n$ identical single barriers into a series of $\nu$ groups consisting of $n$ barriers. The simplest example is given by a quadruple barrier that can be decomposed into two double barriers. An incoming plane wave with a chemical potential $\mu_{n=2,l=1}$ that is in resonance with the double barriers remains an incoming plane wave after passing the first double barrier so that it can also pass the second double barrier in the same manner.
Thus the quadruple barrier is transparent at $\mu_{n=2,l=1}=\mu_{n=4,l=2}$. This argument still holds in the case of a finite interaction strength $g \ne 0$.

By means of a Taylor expansion of the denominator in equation (\ref{Tq_lin}) the transmission coefficient in the vicinity of a resonance with chemical potential $\mu=\mu_{n,l}$ can be written as a Lorentzian (cf.~e.~g.~\cite{Yama90})
\be
   |T_n(\mu)|^2\approx \left[1+\frac{(\mu-\mu_{n,l})^2}{\Gamma_{n,l}^2/4} \right]^{-1}=\frac{\Gamma_{n,l}^2/4}{(\mu-\mu_{n,l})^2+\Gamma_{n,l}^2/4}
\ee
where 
\begin{eqnarray}
  \Gamma_{n,l}&=& 2 \left[\frac{\lambda}{k}\frac{\rd U_{n-1}(z)}{\rd z} \frac{\rd z}{\rd \mu} \right]^{-1} \Bigg|_{\mu=\mu_{n,l}} \\
 &=&2 \sin^2(l \pi/n)\left[\frac{\lambda}{k}\Big((n+2)zU_n-(n+1)U_{n+1} \Big) \frac{\rd z}{\rd \mu} \right]^{-1} \Bigg|_{\mu=\mu_{n,l}}
\end{eqnarray}
is the full width of the peak at half maximum. The factor $\sin^2(l \pi/n)$, which  varies more strongly in dependence on $l$ than the term in the brackets, indicates that within a group of resonances the peaks in the middle are broader than the peaks at the sides. 

Figure \ref{fig-linT2_5} shows the transmission coefficients for $n=2,3,4,5$ for the potential (\ref{delta_series}) with $\lambda=10$, $d=2$ where units with $\hbar=m=1$ are used as in all figures and numerical calculations in this paper. Only the first two groups of resonances are shown. It can be verified that the positions of the two lowest resonance peaks of $|T_2|^2$ coincide with the positions of the second and fifth resonance peak respectively of $|T_4|^2$ as predicted above.

The groups of resonances and the regions of (almost) zero transmittivity in between correspond to the energy bands and band gaps of an infinitely extended delta-comb (or Kronig-Penney) potential, respectively (see e.~g.~\cite{Grif92} and references therein).
%-------------------------------------------------------------------------------------------------------
\section{Transfer map approach}
\label{sec-Multi_transfer}
Because of the nonlinearity of the GPE the transmission coefficient in the interacting case $g \ne 0$ can no longer be obtained by the transfer matrix technique. Instead we introduce a method which we call the {\it transfer map} approach.
To this end we make use of an amplitude phase decomposition  
\be
   \psi(x)=\sqrt{S(x)}\exp(\ri \Phi(x)) 
\ee
which yields the relation 
\be
  \Phi'(x)=\frac{j_\rt m}{\hbar S(x)}
\ee
between the density $S(x)$, the derivative of the phsae $\Phi'(x)$ and the density of the total probability current $j_\rt$.
For given values of $j_\rt$ and $\mu$ the {transfer map} is supposed to map the density $S({\tilde x})$ and its derivative $S'({\tilde x})$ given at some position $\tilde x$ on the right hand side of the barrier region onto the corresponding quantities  $S({x})$ and $S'({x})$ at some position $x$ on the left hand side of the barrier region.

In order to obtain the transfer map of the potential (\ref{delta_series}) we first consider the case 
of a constant potential $V(x)=V_0$ in which an analytical solution for the density $S(x)$ is given by
(cf.~\cite{Carr00a,Carr00b}) 
\be
   S(x)=\varepsilon + \varphi \dn^2(\varrho x +\delta|p) \, . %\qquad \text{ and } \qquad \Phi'(x)=\frac{j_\rt m}{\hbar S(x)} 
\label{S_multi}
\ee
Using the abbreviation $u=\varrho x +\delta$ the derivative of $S(x)$ and its square are given by
\be
   S'(x)=-2 \varrho \,  \varphi \,  p  \, \sn(u|p)\, \cn(u|p)\, \dn(u|p)
   \label{Sp_multi}
\ee
and, by means of the addition theorems of the Jacobian elliptic functions \cite{Abra72},
\be
   S'^2(x)=4 \varphi^2 \varrho^2 \left[(p-1)\dn^2(u|p)+(2-p)\dn^4(u|p)-\dn^6(u|p) \right] \, .
   \label{Sq_prime_multi}
\ee
The parameters in (\ref{S_multi}) must satisfy 
\be
   \varrho^2=-g m\varphi/\hbar^2 \label{rhoQ_multi}
\ee
\be
   \mu-V_0=\frac{3}{2}g \varepsilon +\frac{1}{2}g \varphi (2-p) \label{mu_V0_multi}
\ee
\be
   m j_\rt+ (p-1) g \varphi^2 \varepsilon  - 2(\mu-V_0) \varepsilon^2 + 2 g \varepsilon^3 =0 \, .
   \label{mu_jt}
\ee
Equation (\ref{mu_V0_multi}) can be rewritten as
\be
   \varphi^2(p-1)=\varphi^2 +\left(3 \varepsilon -\frac{2(\mu-V_0)}{g} \right) \varphi \, .
   \label{varphiQ}
\ee
%$j_\rt=\hbar S(x)\phi'(x)/m$
Combining equations (\ref{varphiQ}) and (\ref{mu_jt}) we obtain a quadratic equation for $\varphi$
\be
  \varphi^2+\left(3 \varepsilon- \frac{2(\mu-V_0)}{g}\right)\varphi +\frac{m j_\rt^2}{g \varepsilon}+2 \varepsilon^2- \frac{2(\mu-V_0)}{g}\varepsilon =0
\ee
with the solutions
\be
   \fl \varphi_\pm = -\left( \frac{3}{2}\varepsilon-\frac{\mu-V_0}{g}\right)\pm \sqrt{\left( \frac{3}{2}\varepsilon-\frac{\mu-V_0}{g}\right)^2-2 \varepsilon^2+\frac{2(\mu-V_0)}{g}\varepsilon-\frac{m j_\rt^2}{g \varepsilon}} \label{phi_pm_multi}
\ee
for $g > 0\, (+)$ and $g>0 \, (-)$, respectively.
Suppose the values of $S(\tx)$ and $S'(\tx)$ are known at some position $x=\tx$. The the Jacobian elliptic function $\dn(\tu|p)$ can be expressed as
\be
   \dn^2(\tu|p)=(S(\tx)-\varepsilon)/\varphi \label{dnQ_multi}
\ee
with $\tu =\varrho \tx+\delta$. Thus equation (\ref{Sq_prime_multi}) at $x=\tx$ can be written as
\be
  \frac{-\hbar^2 S'(\tx)^2}{4gm}=\varphi^2(p-1)(S(\tx)-\varepsilon)+(2-p)\varphi(S(\tx)-\varepsilon)^2-(S(\tx)-\varepsilon)^3 \, . \label{Sq_prime_multi2}
\ee
Using equation (\ref{mu_V0_multi}) to eliminate $\varphi$ in (\ref{Sq_prime_multi2}) we finally arrive at the cubic equation
\be
   \fl \varepsilon^3-\frac{2(\mu-V_0)}{g} \varepsilon^2+\left(\frac{\hbar^2 S'^2(\tx)}{4 g m S(\tx)}+\frac{m j_\rt^2}{g S(\tx)}+\frac{2(\mu-V_0)}{g}S(\tx)-S^2(\tx) \right) \varepsilon -\frac{m j_\rt^2}{g}=0 \, ,
   \label{ep_cube}
\ee
the real solution of which is $\varepsilon$. Now that we know $\varepsilon$ the value of $\varphi$ follows from equation (\ref{phi_pm_multi}). From equations (\ref{rhoQ_multi}) and (\ref{mu_V0_multi}) we furthermore obtain
\be
   \varrho=\sqrt{-gm\varphi}/\hbar \quad \quad
\text{ and } \quad \quad
   p=2-\frac{2(\mu-V_0)-3 g \varepsilon}{g \varphi} \, \label{p_multi}.
\ee

Now the only quantity that remains to be computed is the phase $\delta$  of the Jacobi elliptic function.
This can be obtained by numerically solving equation (\ref{dnQ_multi}) at $x=\tx$. However it is more efficient to use the addition theorems
\begin{eqnarray}
\sn(\varrho x +\delta|p)&=&\frac{\sn(v|p)\cn(\tu|p)\dn(\tu|p)+\sn(\tu|p)\cn(v|p)\dn(v|p)}{1-p \, \sn^2(\tu|p)\sn^2(v|p)} \quad \quad \quad  \\
\cn(\varrho x +\delta|p)&=&\frac{\cn(v|p)\cn(\tu|p)-\sn(v|p)\dn(v|p)\sn(\tu|p)\dn(\tu|p)}{1-p \, \sn^2(\tu|p)\sn^2(v|p)}\\
\dn(\varrho x +\delta|p)&=&\frac{\dn(v|p)\dn(\tu|p)-p\,\sn(v|p)\cn(v|p)\sn(\tu|p)\cn(\tu|p)}{1-p \, \sn^2(\tu|p)\sn^2(v|p)}
\end{eqnarray}
instead, where $v=\varrho (x-\tx)$ and $\tu =\varrho \tx+\delta$. In order to apply these addition theorems the values of the Jacobian elliptic functions at $x=\tx$ are required. The Jacobian function $\dn(\tu|p)$ is given by equation (\ref{dnQ_multi}) and the remaining functions can be computed via $\cn(\tu|p)=\cos({\rm am}\, \tu)$ and $\sn(\tu|p)=\sin({\rm am}\, \tu)$ where
\be
   {\rm am}\, \tu = \pm \arcsin \left(\sqrt{1-\dn^2(\tu|p)/p} \right)
   \label{amu}
\ee
is the so-called \emph{amplitude} of the Jacobian elliptic functions. The sign must be chosen such that $\sgn\left(\sin({\rm am}\, \tu)\cos({\rm am}\,\tu)\right)=\sgn \left(-S'(\tx)/\varphi \right)=\sgn(\sn(\tu|p)\cn(\tu|p)\dn(\tu|p))$.
Thus for given values of $\mu-V_0$ and $j_\rt$ equations (\ref{S_multi}), (\ref{Sp_multi}), (\ref{phi_pm_multi}) and (\ref{ep_cube})-(\ref{amu}) define a map
\be
    {\cal U}_{\mu-V_0,j_t,\tx-x} \quad : \quad \left(S(\tx),S'(\tx) \right) \longmapsto \left(S(x),S'(x) \right)
    \label{transfer_constant}
\ee
which we call the transfer map of the constant potential $V(x)=V_0$.
% \be
%    \cal{D}_{\lambda,x_0}: \left(S(x_0+0),S'(x_+0) \right) \longmapsto  \left(S(x_0-0)=S(x_0+0),S'(x_0-0)=S'(x_+0)-4 \lambda S(x_0+0) \right)
% \ee
% \be
%    {\cal D}_{\lambda}: \left(S(x+0),S'(x+0) \right) \longmapsto  \left(S(x-0)=S(x+0),S'(x-0)=S'(x+0)-4 \lambda S(x+0) \right)
% \ee
The matching condition for the wavefunction $\psi(x)$ and its derivative $\psi'(x)$ at the position $x_0$ of a delta potential with the strength $\hbar^2 \lambda/m$ is given by $\psi'(x_0-)=\psi(x_0+)$ and $\psi'(x_0-)=\psi'(x_0+)- 2 \lambda \psi(x_0)$.
Straightforward algebra shows that these conditions read
\be
    S(x_0-)=S(x_0+) \text{ and } S'(x_0-)=S'(x_0+)- 4 \lambda S(x_0) \label{4lambda}
\ee
in terms of $S(x)=|\psi(x)|^2$ and its derivative. This leads to the map
\be
   {\cal D}_{\lambda} \quad:\quad \left(S,S' \right) \longmapsto  \left(S,S'-4 \lambda S \right)
\ee
for the delta potential with strength $\hbar^2 \lambda/m$. Thus we obtain the transfer map
\be
   {\cal M}_{n}=\left({\cal D}_{\lambda} {\cal U}_{\mu,j_t,d}\right)^{n-1}{\cal D}_{\lambda}
   \label{transfer_multi}
\ee
of the potential (\ref{delta_series}) with $V_0=0$.

For a stationary scattering state the solution in the region $x>(n-1)d$, i.e.~on the right hand side of the potential barriers, is given by a plane wave $C \exp(\ri k_\rC x)$ with $k_\rC=\sqrt{2m(\mu-g|C|^2)/\hbar}$ and the total current can be expressed as
\be
   j_\rt=|C|^2 \hbar k_\rC /m \, . \label{j_t_multi}
\ee
The probability density and its derivative at the right hand side of the barriers are $ \quad S((n-1)d+)=|C|^2$ and $S'((n-1)d+)=0$. The transfer map (\ref{transfer_multi}) determines the probability density and its derivative at the left hand side of the barriers via
\be
   \left(S(0-),S'(0-) \right)={\cal M}_{n}\left(|C|^2,0\right) \,. \label{transfer_multi2}
\ee
For the parameters considered in this paper the mean--field interaction potential $g S(x)$ outside the potential is negligibly small compared to the chemical potential $\mu$ so that we can write the wavefunction in the region $x<0$ as a superposition $A \exp(\ri kx)+ b \exp(-\ri kx)$ of an incoming and an outgoing plane wave. For the wavefunction $\psi(0-)$ and its derivative $\psi'(0-$ at $x=0$ we thus obtain the condition
\be
    2 \ri k A=\psi'(0-)+\ri k \psi(0-)\, , \label{2ikA_multi}
\ee
with the incoming wave amplitude $A$ and the wavenumber $k=\sqrt{2m \mu}/\hbar$.
In order to express this condition in terms of $S(0-)$ and $S'(0-)$ we multiply it by $\psi^*(0-)$  arriving at
\be
    2 \ri k \psi^*(0-) A=\psi^* \psi'(0-)+\ri k |\psi(0-)|^2 \label{2ikA_multi2} \, .
\ee
Using $S'(0-)=\psi^*(0-)\psi'(0-)+\psi^{*'}(0-)\psi(0-)$ and $$j_\rt=-\ri \hbar \big( \psi^*(0-)\psi'(0-)-\psi^{*'}(0-)\psi(0-)\big)/(2m)$$ to replace the term $\psi^*(0-)\psi'(0-)$ in equation (\ref{2ikA_multi2}) we obtain
\be
    2 \ri k \psi^*(0-) A=S'(0-)/2 +\ri \left(k_\rC |C|^2+ k S(0-)\right) \label{2ikA_multi3}
\ee
where we have used equation (\ref{j_t_multi}) to replace $j_\rt$.
The absolute square of equation (\ref{2ikA_multi3}) provides a convenient condition
\be
    4 k^2 S(0-)^2 |A|^2=S'^2(0-)/4+\left(k_\rC |C|^2+ k S(0-) \right)^2 \label{Multi_cond}
\ee
for the wavefunction in the upstream region. 
For the parameter range considered in this paper the approximation $k_\rC \approx k=\sqrt{2m\mu}/\hbar$ can be made since the effective nonlinearity is small outside the barrier region (cf.~the discussion above).

Numerically, for given values of $\mu$ and $A$ the squared magnitude $|C|^2$ of the outgoing wave amplitude is obtained by solving the system (\ref{transfer_multi2}), (\ref{Multi_cond}). This is achieved by combining a bisection method with a finite grid for $|C|^2$. 
The transmission coefficient is then given by $|T|^2=j_\rt/j_{\rm in}\approx |C|^2/|A|^2$. %k_\rC |C|^2/\left(k |A|^2\right)$.

The transfer map (\ref{transfer_constant}) of the constant potential $V(x)=V_0$ simplifies considerably in the special case $S'(\tx)=0$ in which equation (\ref{ep_cube}) reads
\be
    (\varepsilon-S(\tx))\left[\varepsilon^2+ \left(S(\tx)-\frac{2(\mu-V_0)}{g}\right)\varepsilon+\frac{m j_t^2}{g S(\tx)} \right]=0 \, .
\ee
Apart from the  trivial solution $S(x)=\varepsilon=const$, $\varphi=0$ this equation has the solutions
\be
   \varepsilon_\pm=\left(\frac{\mu-V_0}{g}-\frac{S(\tx)}{2} \right)\pm \sqrt{\left(\frac{\mu-V_0}{g}-\frac{S(\tx)}{2} \right)^2-\frac{m j_t^2}{g S(\tx)}}
\ee
for $g>0\, (+)$ and $g<0\, (-)$ respectively.
For $g<0$ equation (\ref{Sp_multi}) leads to (cf.~\cite{06nl_transport})
\be
   \varphi=S(\tx)-\epsilon \, ,
\ee
for $g>0$ it yields
$
   S(\tx)=\varepsilon+\varphi(1-p)
$
and finally, together with equation (\ref{p_multi}),
\be
   \varphi=\frac{2(\mu-V_0)}{g}-2 \varepsilon-S(\tx) \, .
\ee
The phase shift is given by
\be
   \delta=- \varrho \tx \: (g>0) \text{ or } \delta=K(p)-\varrho \tx (g<0)
\ee
(cf.~\cite{06nl_transport}).

In the following section the transfer map approach is applied to double, triple, quadruple and quintuple barrier tunnelling.
%--------------------------------------------------------------------------------------------------
\section{Nonlinear multi--barrier transmission}
\label{sec-Multi_Tq}
\subsection{Double barrier}

\begin{figure}[htb]
\centering
\includegraphics[width=7.5cm,  angle=0]{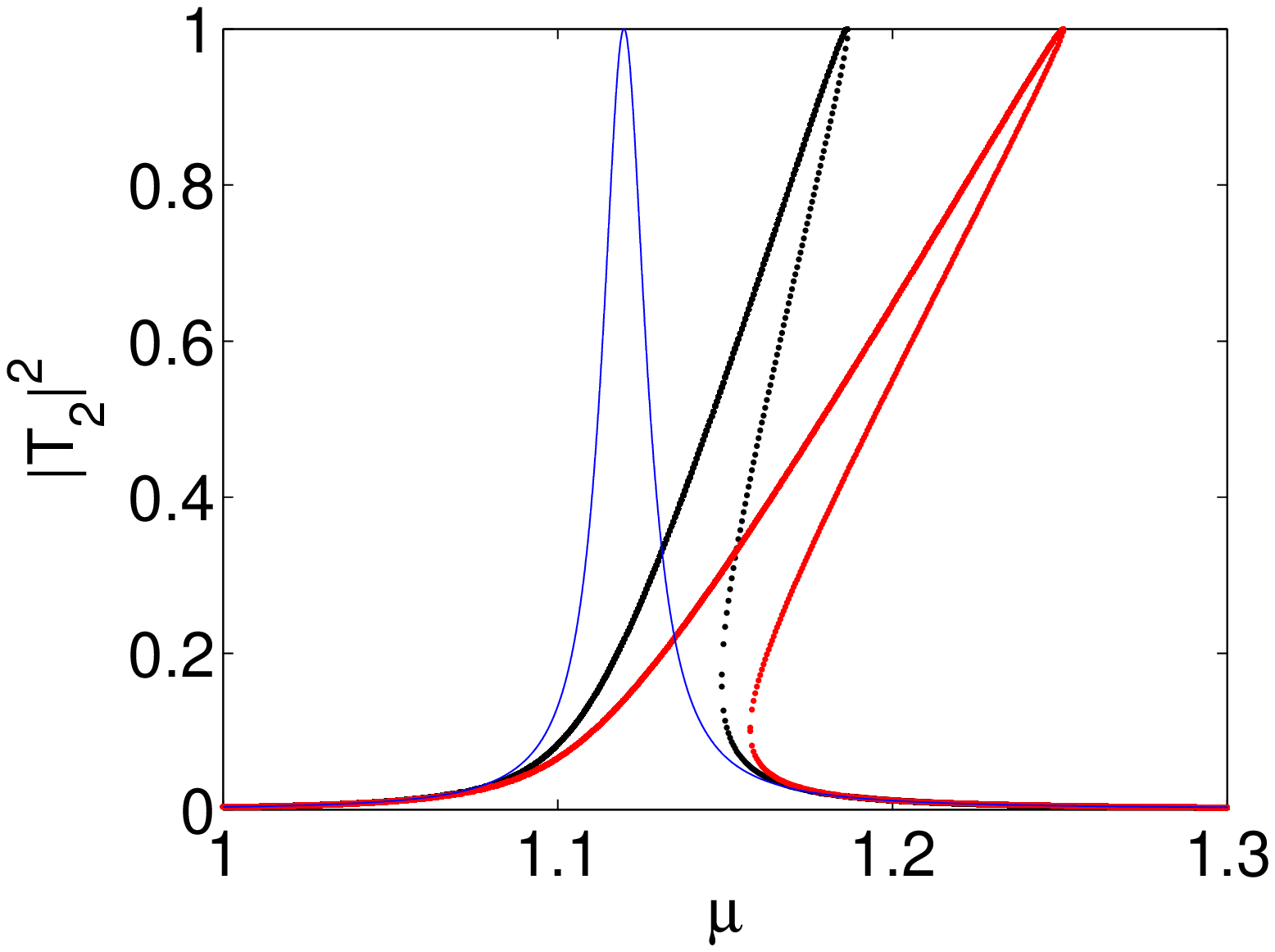}
\includegraphics[width=7.5cm,  angle=0]{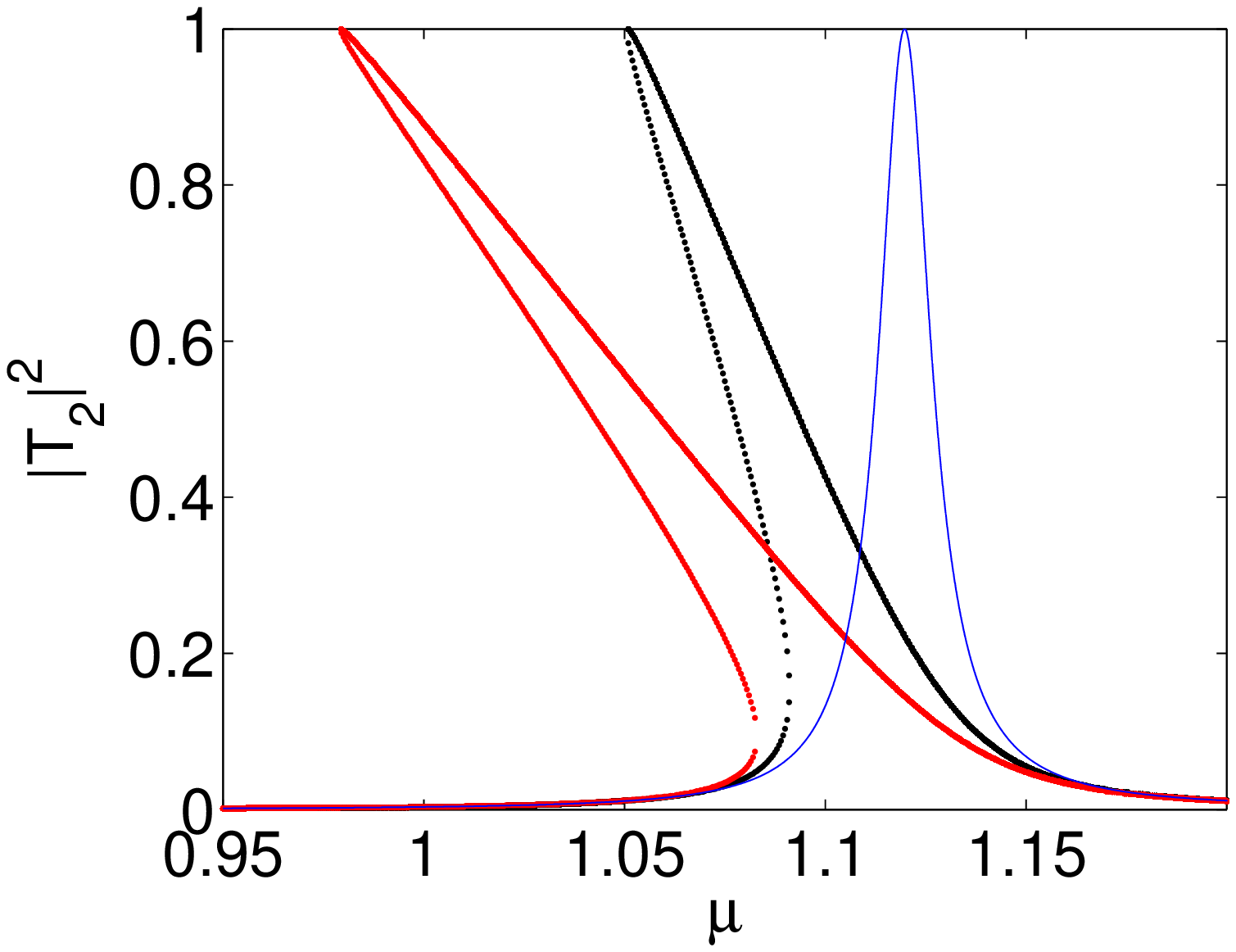}\\
\caption{\label{fig-posT2} {Transmission coefficient $|T_2|^2$ in dependence on the chemical potential $\mu$ for $\lambda=10$, $d=2$, $A=0.1$. Left panel: $g=0$ (solid blue line), $g=0.005$ (black dots), $g=0.1$ (red dots). Right panel: $g=0$ (solid blue line), $g=-0.05$ (black dots), $g=-0.1$ (red dots).
}}
\end{figure}
Using the transfer map approach described in the previous section we compute the transmission coefficient $|T_2|^2$ in dependence on the chemical potential $\mu$ for the potential (\ref{delta_series}) with $n=2$ barriers, potential strength $\lambda=10$, distance $d=2$ (cf.~figure \ref{fig-linT2_5}) and an incoming amplitude $A=0.1$ which is displayed in figure \ref{fig-posT2} for several values of the interaction parameter $g$. We obtain the familiar behaviour for nonlinear single well/double barrier tunnelling (see e.~g.~\cite{Paul05,06nl_transport,Paul07b,08nlLorentz}):
For $g>0$, the peaks are shifted to higher chemical potentials due to the repulsive mean--field term $g|\psi(x)|^2$ in the GPE. The wavefunctions of the linear ($g=0$) system corresponding to resonant transmission $|T|^2 \approx 1$ are more strongly affected by the mean--field term  than those corresponding to off-resonant transmission because they have a greater total norm $\int_{0}^d \rd x \,  |\psi(x)|^2$  inside the well. Thus the maximum of a peak experiences a stronger shift than its flanks so that the peak bends more and more to the right for increasing nonlinearity $g$, leading to bistability. Analogously a peak bends to the left for an attractive interaction $g<0$.
In other words, in some parameter regions there exist states with the same chemical potential but with different average numbers of particles inside the well corresponding to different values of the transmission coefficient. The transmission coefficient is thus subject to a hysteresis effect as the system has a memory given by the average number of particles inside the well. As an example we consider a  transmission coefficient for repulsive nonlinearity as shown in the left panel of figure \ref{fig-posT2} for $g=+0.1$ (red). Let us assume that the system is initially prepared in a transmission state corresponding to a chemical potential $\mu\approx 1.1$ on the left hand side of the bistable region. If the chemical potential $\mu$ of the incoming matter wave is slowly increased the values of the transmission coefficient follow the upper curve in figure \ref{fig-posT2} until the end of the bistable region is reached. Then the transmittivity ``drops down'' and follows the only existing branch.
This behaviour has been explicitly demonstrated in a recent numerical study \cite{Erns09} for a double Gaussian barrier using the time--dependent description given in equation (\ref{GPE_t}).

%The wavefunctions of the linear ($g=0$ system) corresponding to off-resonant transmission $|T|^2<1$ are less strongly affected by the nonlinear term $g|\psi(x)|^2$ in the GPE than the resonance wavefunctions because they have a smaller total norm $\int_{-a}^a \rd x \,  |\psi(x)|^2$ in inside the well. Thus the peaks of the transmission coefficient are deformed rather than just shifted to lower chemical potentials.
%In accordance with the results for the transmission through a square well (see chapter \ref{chap-sqw}) the curves bend to the right (left) for an increasingly positive (negative) value of the interaction strength $g$ leading to bistability.

\subsection{Triple barrier}
% \begin{figure}[htb]
% \centering
% \includegraphics[width=7.5cm,  angle=0]{/home/aleph/rapedius/Programme/TransNeu/md_zwischen0.eps}
% \includegraphics[width=7.5cm,  angle=0]{/home/aleph/rapedius/Programme/TransNeu/md_zwischen.eps}\\
% \includegraphics[width=7.5cm,  angle=0]{/home/aleph/rapedius/Programme/TransNeu/md_zwischen1.eps}
% \includegraphics[width=7.5cm,  angle=0]{/home/aleph/rapedius/Programme/TransNeu/md_zwischen2.eps}
% \caption{\label{fig-posT3} {Transmission coefficient $|T_3|^2$ in dependence on the chemical potential $\mu$ for $\lambda=10$, $d=2$, $A=0.1$. Upper left panel: $g=0$ (solid blue line), $g=0.017$ (black dots), $g=0.034$ (red dots). Upper right panel: $g=0.036$ (black dots), $g=0.039$ (red dots). Lower left panel: $g=0.05$ (black dots), $g=0.1$ (red dots). Lower right panel: $g=0.25$ (black dots), $g=0.5$ (red dots).}}
% \end{figure}
\begin{figure}[htb]
\centering
\includegraphics[width=7.5cm,  angle=0]{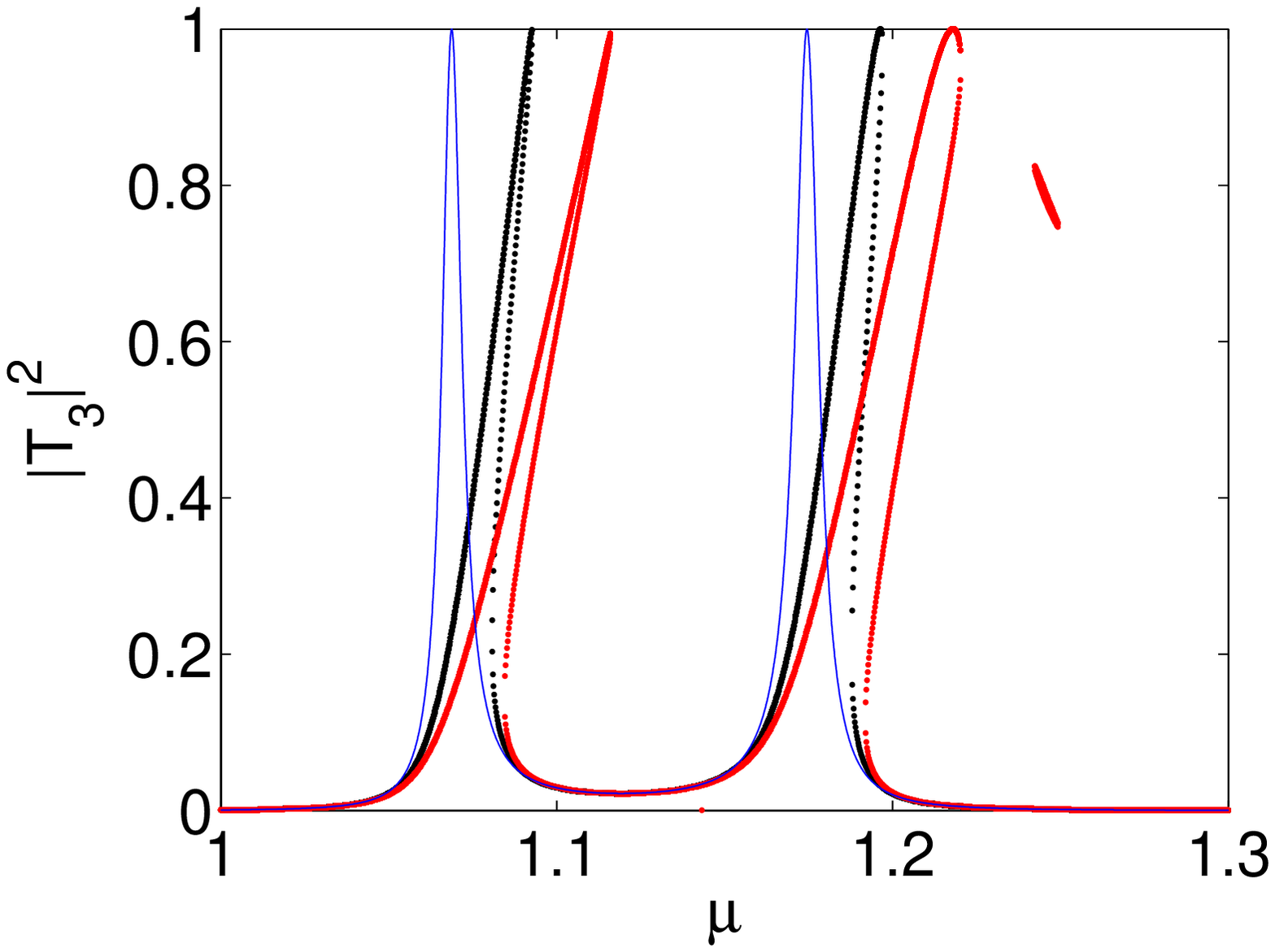}
\includegraphics[width=7.5cm,  angle=0]{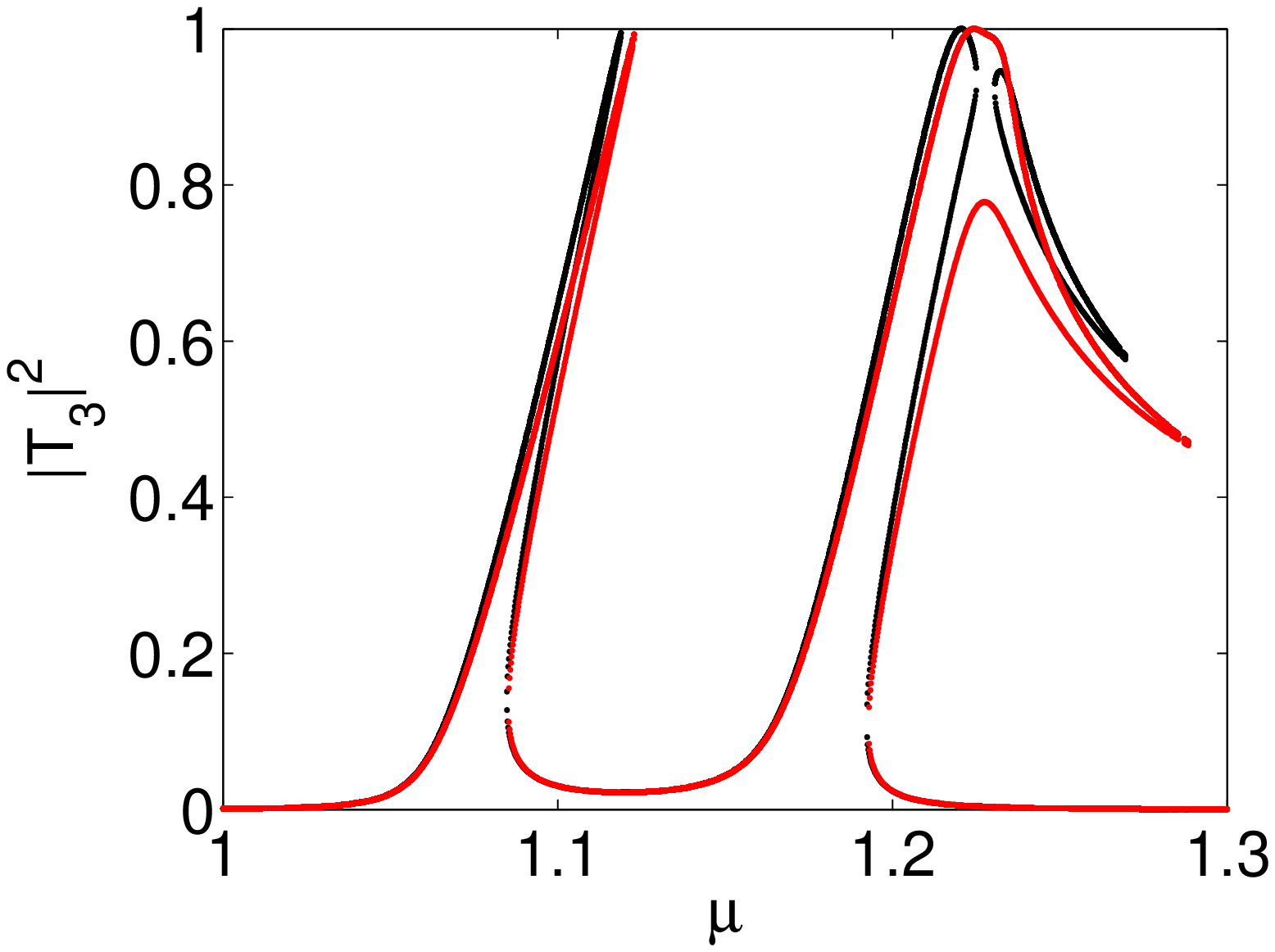}\\
\includegraphics[width=7.5cm,  angle=0]{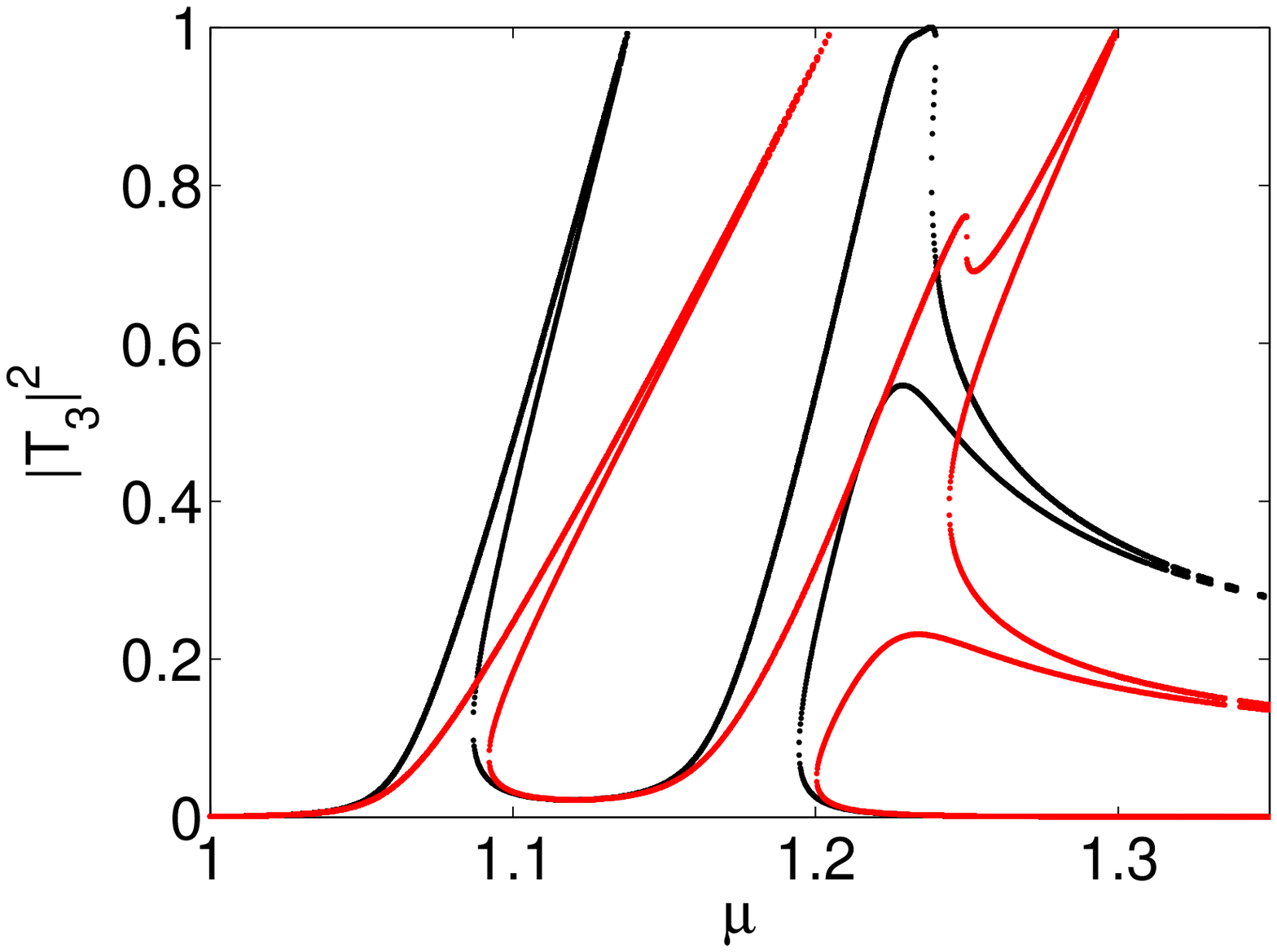}
\includegraphics[width=7.5cm,  angle=0]{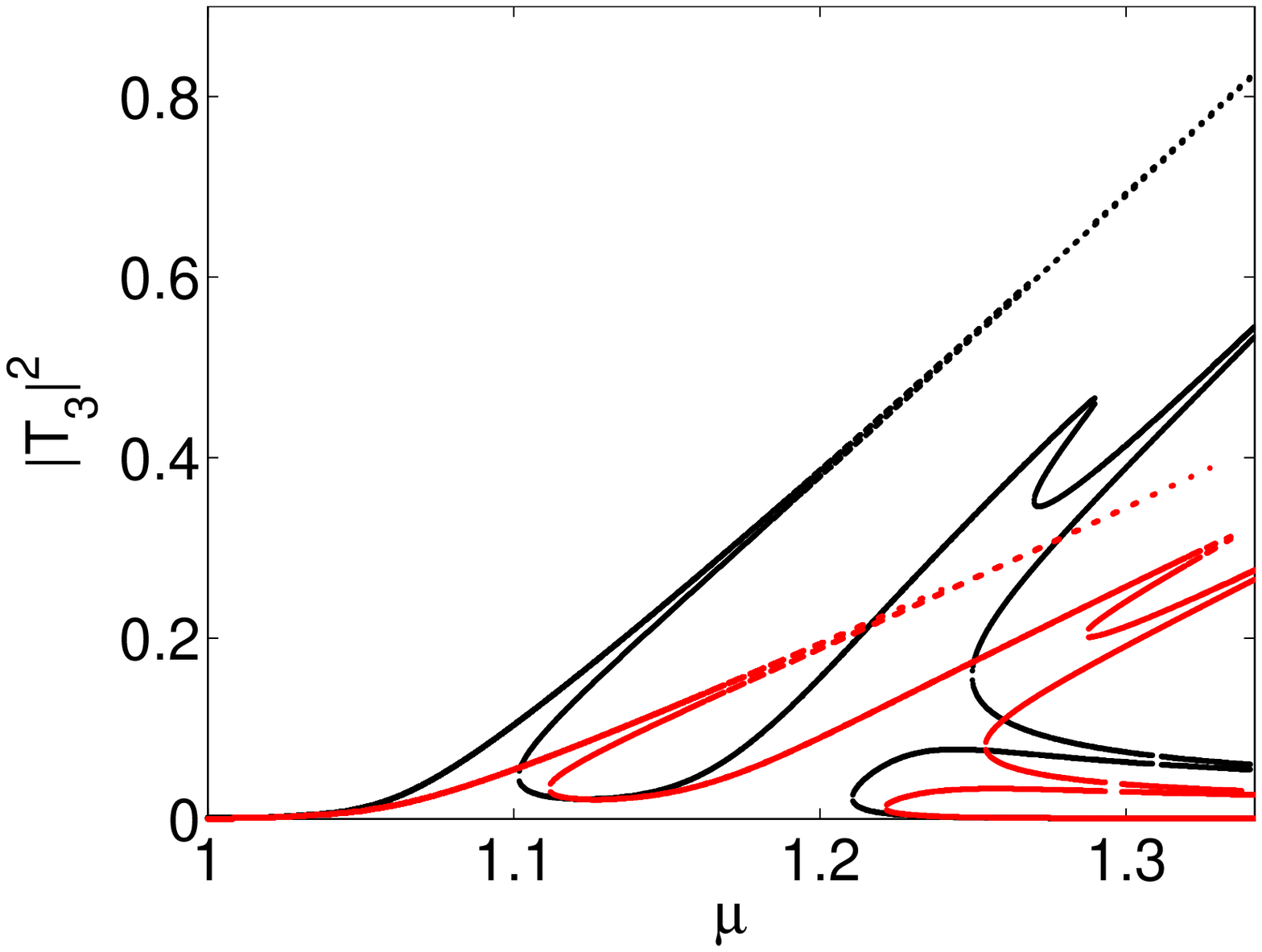}
\caption{\label{fig-posT3} {Transmission coefficient $|T_3|^2$ in dependence on the chemical potential $\mu$ for $\lambda=10$, $d=2$, $A=0.1$. Upper left panel: $g=0$ (solid blue line), $g=0.017$ (black dots), $g=0.034$ (red dots). Upper right panel: $g=0.036$ (black dots), $g=0.039$ (red dots). Lower left panel: $g=0.05$ (black dots), $g=0.1$ (red dots). Lower right panel: $g=0.25$ (black dots), $g=0.5$ (red dots).}}
\end{figure}
\begin{figure}[ht]
\centering
\includegraphics[width=10cm,  angle=0]{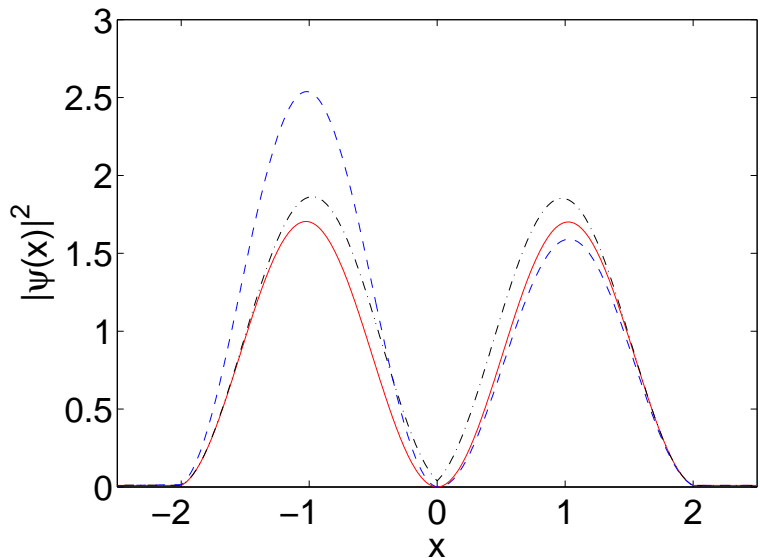}
\caption{\label{fig-WF3} {Squared magnitudes $|\psi(x)|^2$ of wavefunctions corresponding the peak maxima in the transmission coefficient of the triple barrier with $\lambda=10$, $d=2$, $A=0.1$ and nonlinearity $g=0.036$ (cf.~the black curve in the upper right panel of figure \ref{fig-posT3}). Solid red: first maximum with $\mu \approx 1.12$, $|T|^2 \approx 1$. Dashed dotted black: second maximum with $\mu \approx 1.22$, $|T|^2 \approx 1$.
Dashed blue: maximum of the looped structure with $\mu \approx 1.23$, $|T|^2 \approx 0.95$.}}
\end{figure}
\begin{figure}[ht]
\centering
\includegraphics[width=7.5cm,  angle=0]{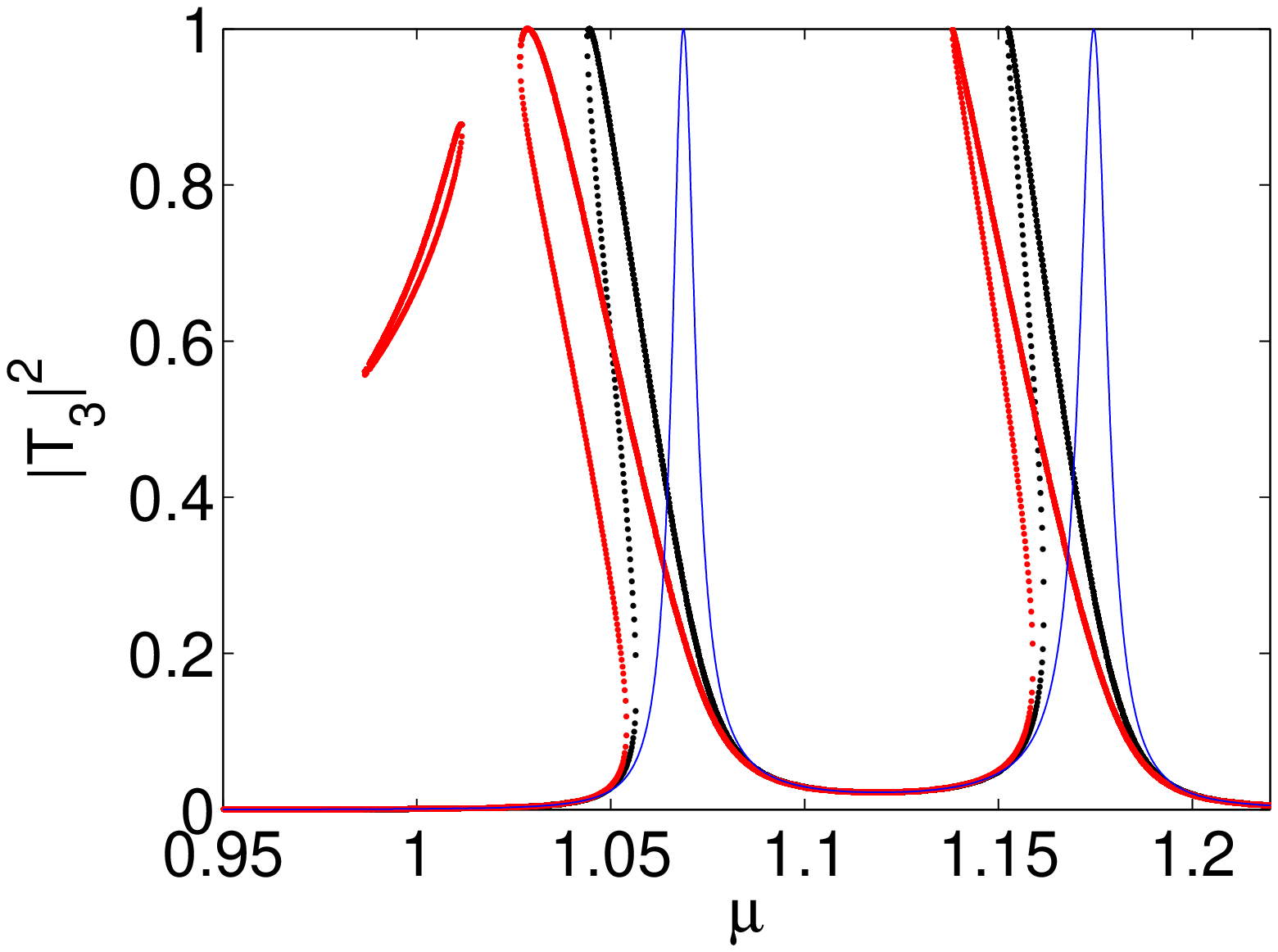}
\includegraphics[width=7.5cm,  angle=0]{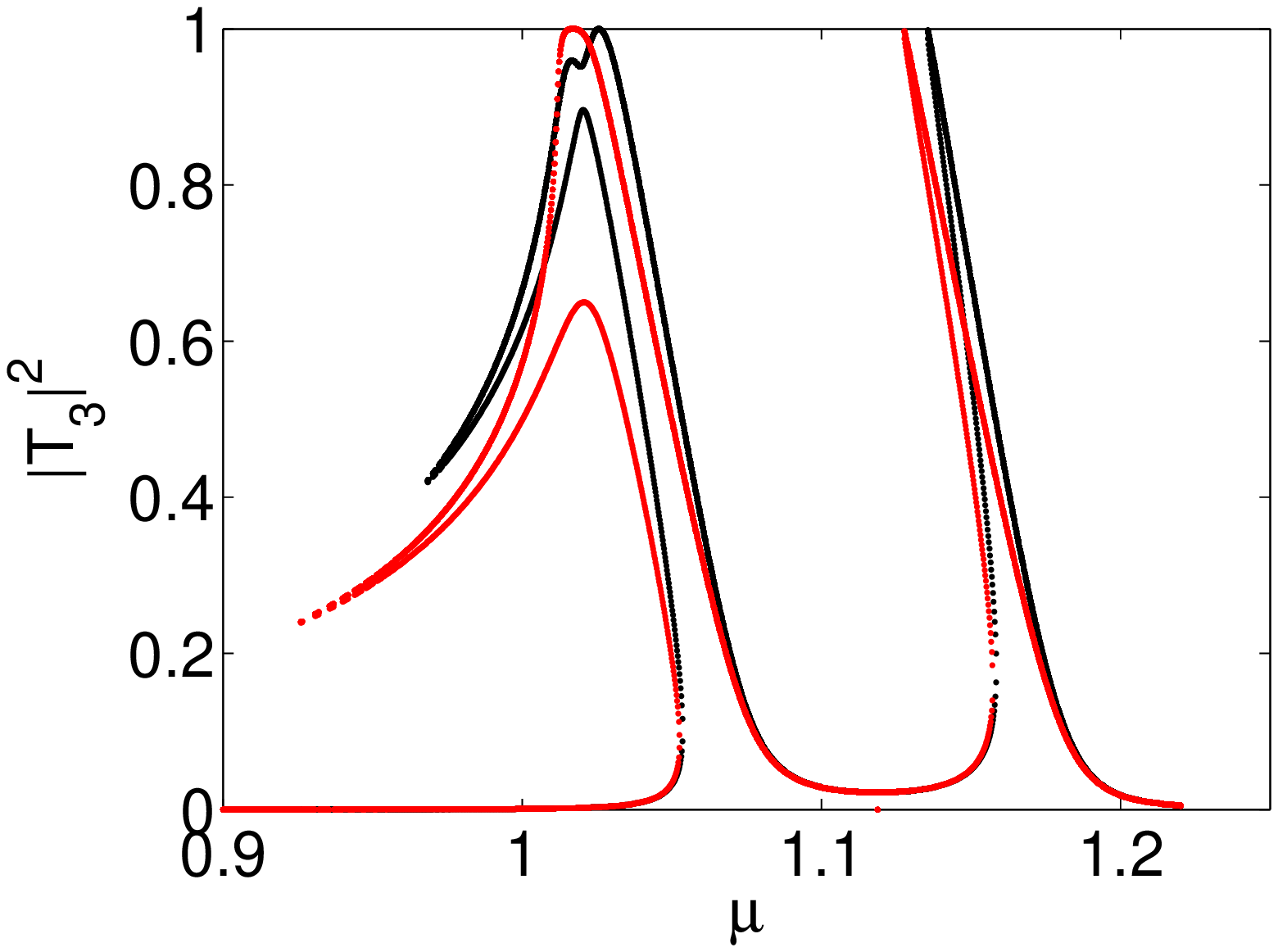}\\
\includegraphics[width=7.5cm,  angle=0]{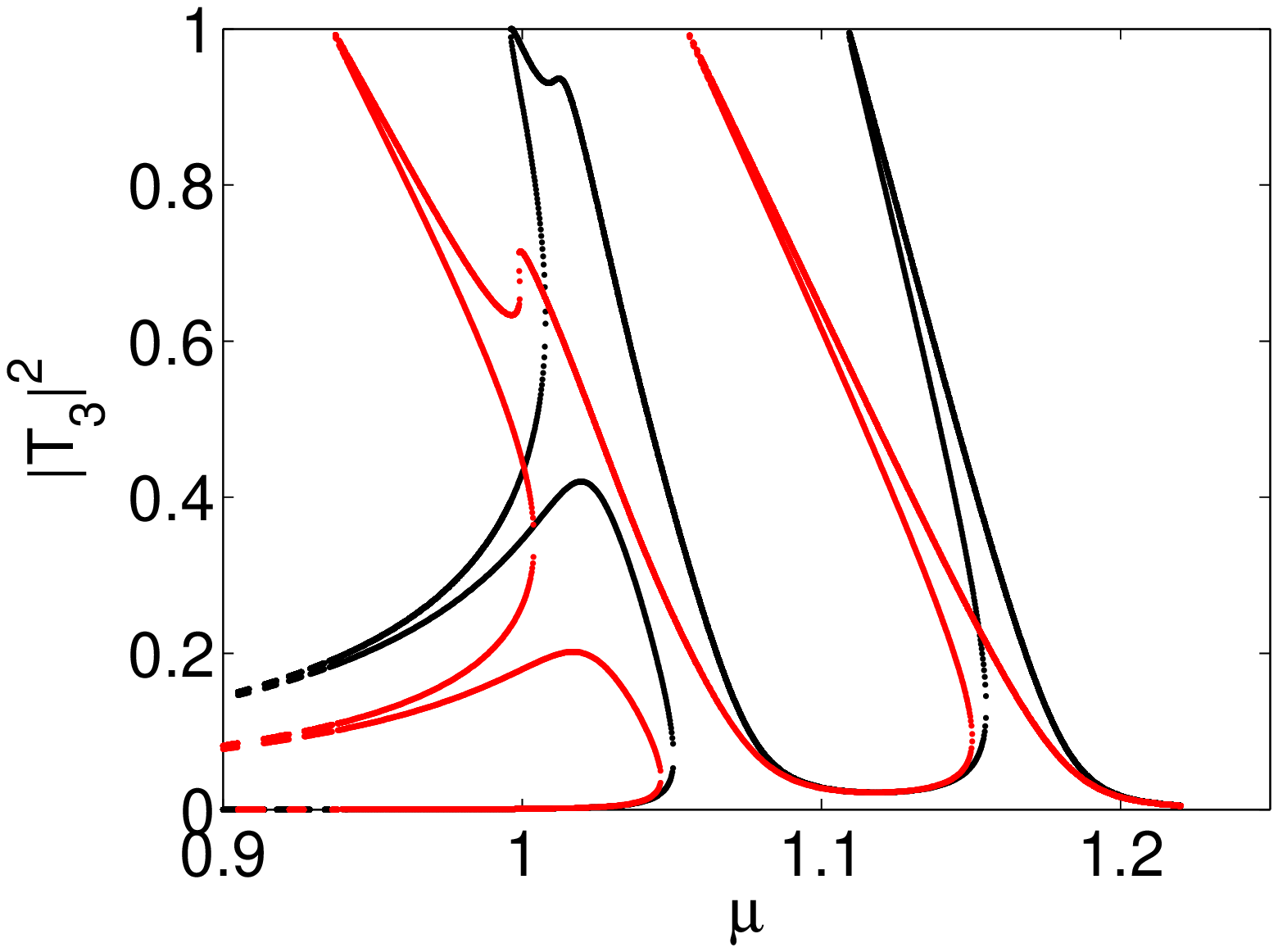}
\includegraphics[width=7.5cm,  angle=0]{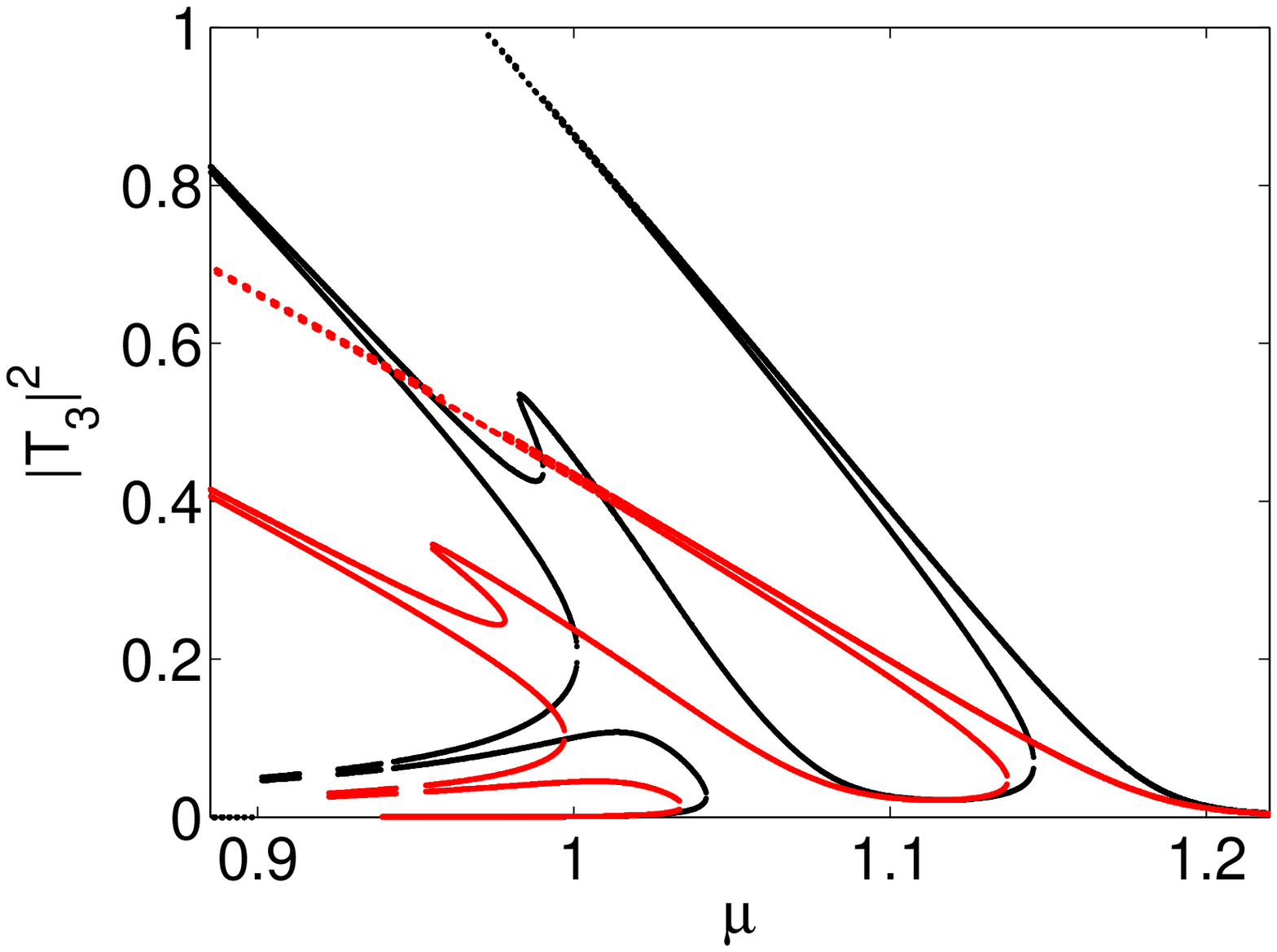}
\caption{\label{fig-negT3} {Transmission coefficient $|T_3|^2$ in dependence on the chemical potential $\mu$ for $\lambda=10$, $d=2$, $A=0.1$. Upper left panel: $g=0$ (solid blue line), $g=-0.017$ (black dots), $g=-0.028$ (red dots). Upper right panel: $g=-0.030$ (black dots), $g=-0.036$ (red dots).
Lower left panel:$g=-0.05$ (black dots), $g=-0.09$ (red dots). Upper right panel: $g=-0.15$ (black dots), $g=-0.3$ (red dots).}}
\end{figure}

Figure \ref{fig-posT3} shows the transmission coefficient $|T_3|^2$ in dependence on the chemical potential $\mu$ for the potential (\ref{delta_series}) with $n=3$ barriers, potential strength $\lambda=10$, distance $L=2$ (cf.~figure \ref{fig-linT2_5}) and an incoming amplitude $A=0.1$ for various positive values of the interaction parameter $g$. For a moderately repulsive $g$ the resonance peaks bend to the right as known for the double barrier. This effect is slightly weaker for the second peak with larger chemical potential $\mu$ than for the first one because the respective kinetic energy is higher in comparison with the mean--field interaction energy (cf.~the discussion  in \cite{06nl_transport}).
For $g \approx 0.035$ a narrow loop which is not connected with the other branches of the transmission coefficient emerges close to the second resonance peak. When $g$ is further increased the second resonance and the loop approach each other. During the process the second resonance peak is slightly deformed until the two structures collide and finally unite. A similar behaviour could be observed for the transmission coefficient of the finite square well considered in \cite{06nl_transport}. There, the looped structures originate from bound states which have been destabilized and thus turned into resonances due to repulsive interaction.
Here, the loop is formed by solutions without a linear counterpart (so called allochtonous solutions). 
By comparison with the double barrier case we can identify the emergence of unconnected loops as an effect of interaction between the first two resonances of the system. % which is analogous to symmetry breaking the bifurcation phenomenon observed for the double delta-shell potential in chapter \ref{chap-DDShell}.

Figure \ref{fig-WF3} displays the squared magnitudes $|\psi(x)|^2$ of the wavefunctions corresponding to the peak maxima (respectively the looped structure) in the transmission coefficient of the triple barrier with $\lambda=10$, $d=2$, $A=0.1$ and nonlinearity $g=0.036$ (cf.~ the black curve in the upper right panel of figure \ref{fig-posT3}). The densities $|\psi(x)|^2$ corresponding to the two autochtonous states with maximum transmission (solid red and dashed dotted black) are symmetric whereas the density corresponding to the maximum of the allochtonous loop with $|T|^2<1$ is asymmetric. Hence this state represents an example of symmetry breaking in a nonlinear system similar to, e.~g., the self-trapping states in double-well potentials (see e.~g.~\cite{Theo06,Khom07,09ddshell,Infe06,Albi05}).

For even higher values of $g$ the transmission peaks in figure \ref{fig-posT3} bend more and more to the right and the second resonance peak develops into a fork (double peak). Note that very narrow structures are not always perfectly resolved because our implementation of the transfer map approach uses a finite grid for the outgoing amplitude $|C|^2$ (see section \ref{sec-Multi_transfer}). 

An analogous behaviour can be observed in figure \ref{fig-negT3} which shows the transmission coefficient $|T_3|^2$ for the same potential as in figure \ref{fig-posT3} for negative values of $g$. Now the curves bend to the left and the looped structure appears in the vicinity of the first resonance peak. %This corresponds to the behaviour of the eigenstates of the double delta-shell potential (chapter \ref{chap-DDShell}) where the allochtonous states emerge in the vicinity of the lower respectively higher autochtonous state for attractive respectively repulsive interaction.

%-------------------------------------------------------------------------------------------------------
%\newpage
\subsection{Quadruple barrier}
\begin{figure}[htb]
\centering
\includegraphics[width=7.5cm,  angle=0]{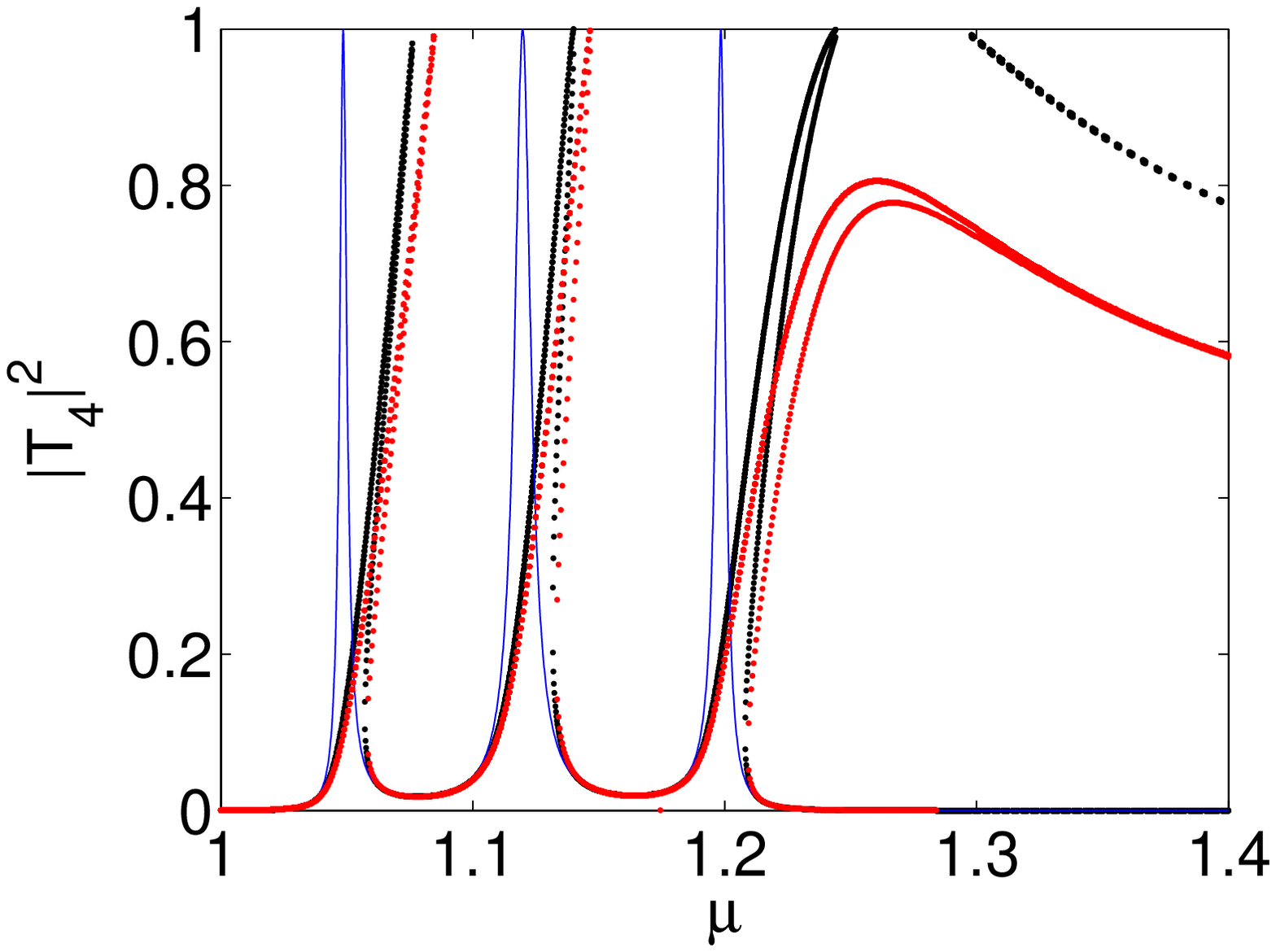}
\includegraphics[width=7.5cm,  angle=0]{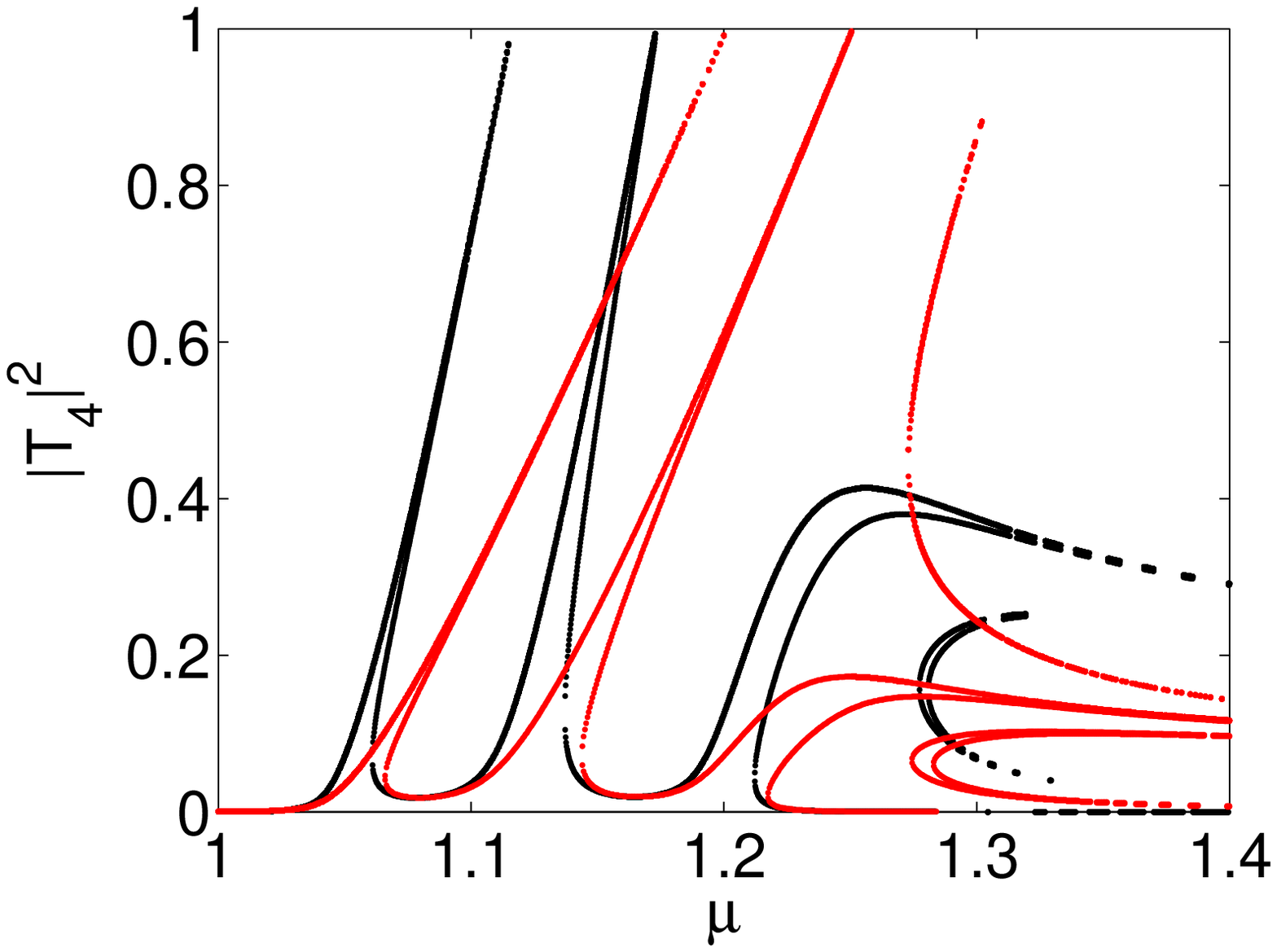}\\
\includegraphics[width=7.5cm,  angle=0]{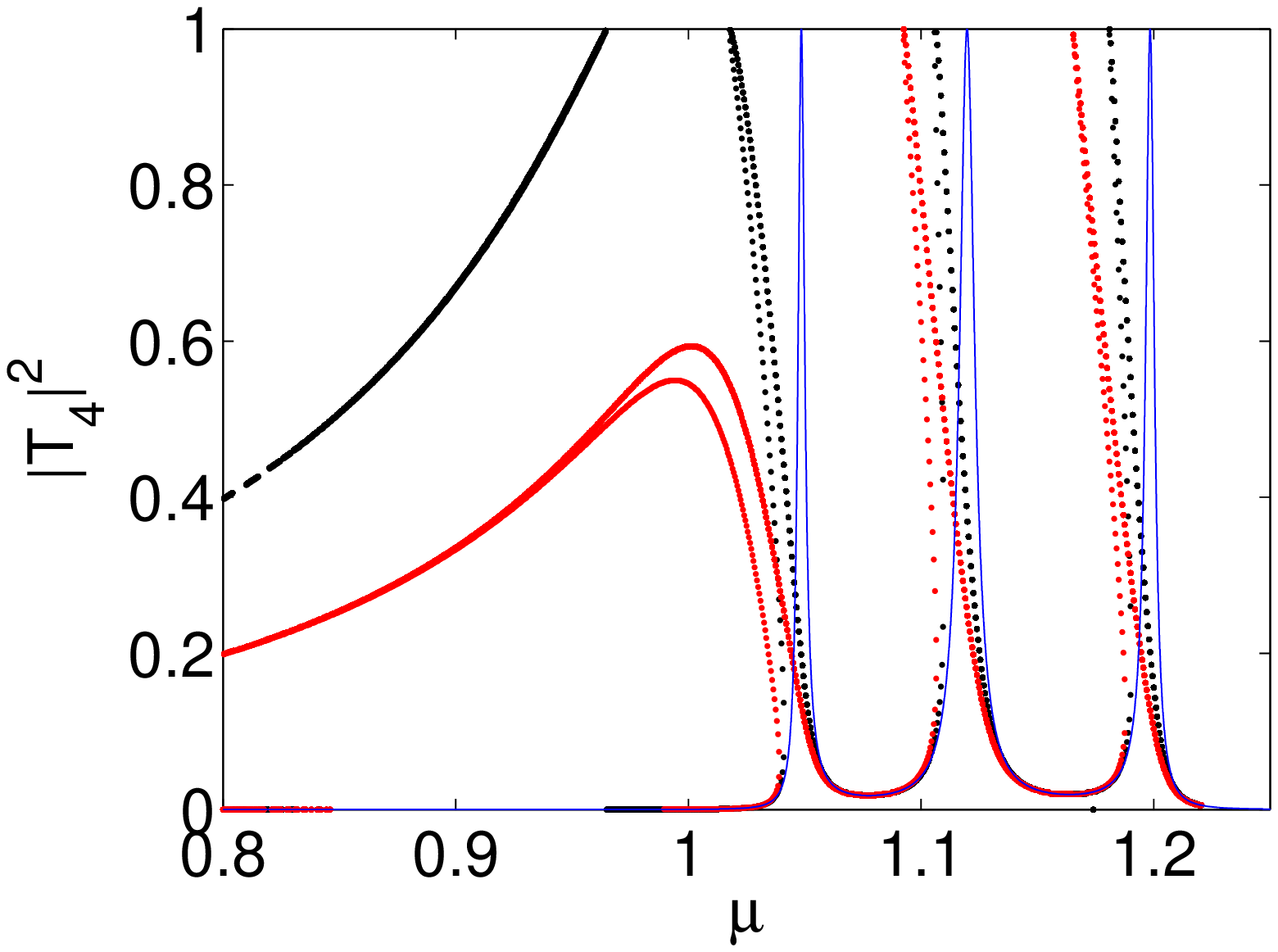}
\includegraphics[width=7.5cm,  angle=0]{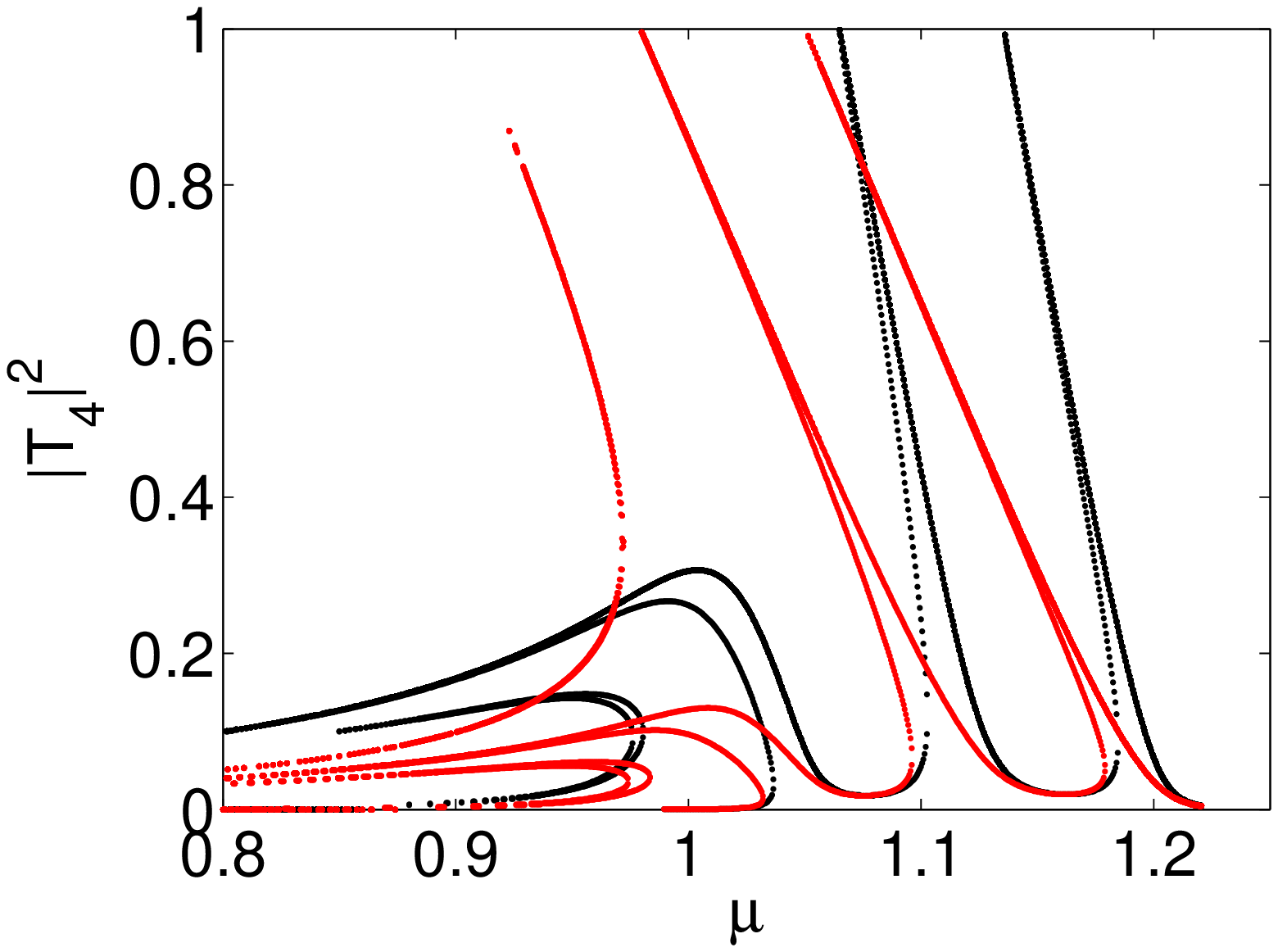}
\caption{\label{fig-posT4} {Transmission coefficient $|T_4|^2$ in dependence on the chemical potential $\mu$ for $\lambda=10$, $d=2$, $A=0.1$. Upper left panel: $g=0$ (solid blue line), $g=0.015$ (black dots), $g=0.02$ (red dots). Upper right panel: $g=0.04$ (black dots), $g=0.1$ (red dots). Lower left panel: $g=0$ (solid blue line), $g=-0.01$ (black dots), $g=-0.02$ (red dots) . Lower right panel:$g=-0.04$ (black dots), $g=-0.1$ (red dots)}}
\end{figure}
Now we consider the transmission coefficient $|T_4|^2$ in dependence on the chemical potential $\mu$ for the potential (\ref{delta_series}) with $n=4$ barriers, $\lambda=10$, $d=2$ and $A=0.1$.
Figure \ref{fig-posT4} displays $|T_4|^2$ for various values of $g$. As in the case of three barriers for positive interaction a narrow structure which is not connected with the other branches of the transmission coefficient emerges at the right hand side of the first group of resonances. In contrast to the triple barrier case, however, the maximum transmittivity within the newly created branch is $|T_4|^2 \approx 1$. For higher values of $g$ the newly created branch unites with the third resonance peak. After the unification the transmittivity in the vicinity of the third resonance peak no longer reaches full transparency which is another difference to the case of three barriers. When $g$ is further increased more and more unconnected branches emerge. As for the triple barrier the situation is completely analogous in the case of negative interaction.
By comparing the respective transmission coefficients for $g=0$ and $g=\pm 0.1$ in figures \ref{fig-posT4} and \ref{fig-posT2} we can directly verify a property predicted in section \ref{sec_Multi_intro}, namely that the position of the second resonance peak of the quadruple barrier always coincides with the position of the first resonance peak of the double barrier.

\subsection{Quintuple barrier}
\begin{figure}[htb]
\centering
\includegraphics[width=7.5cm,  angle=0]{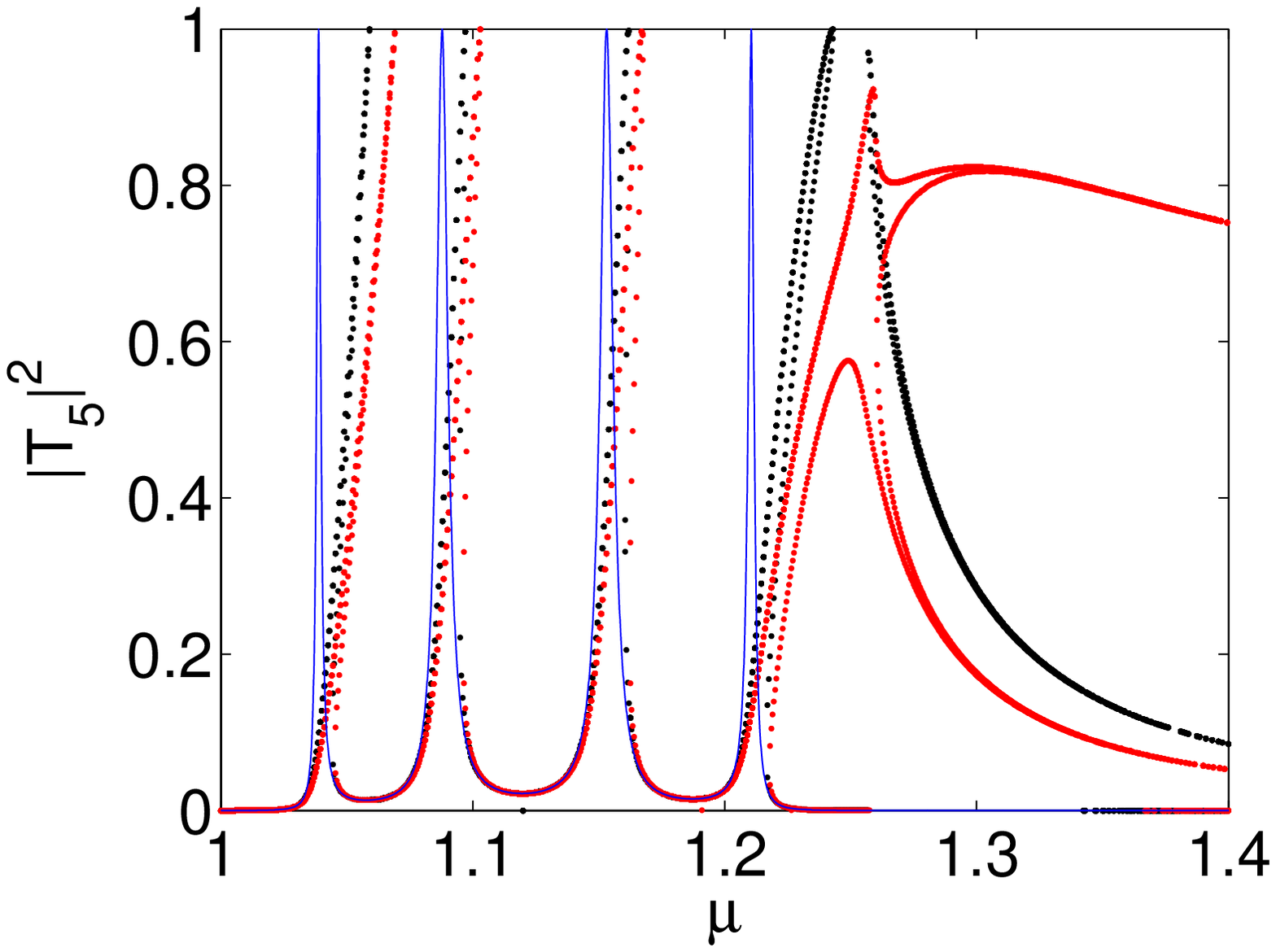}
\includegraphics[width=7.5cm,  angle=0]{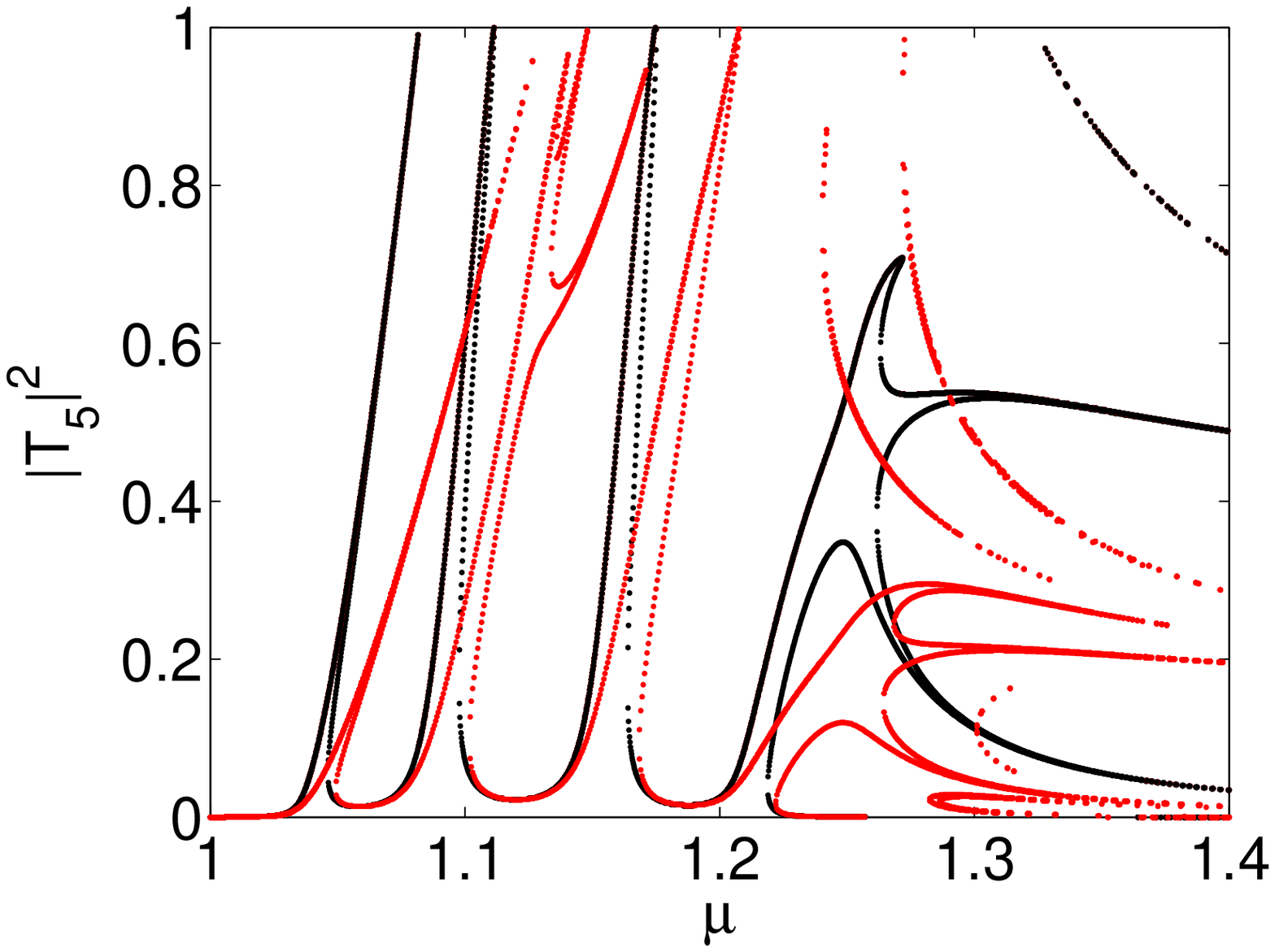}\\
\includegraphics[width=7.5cm,  angle=0]{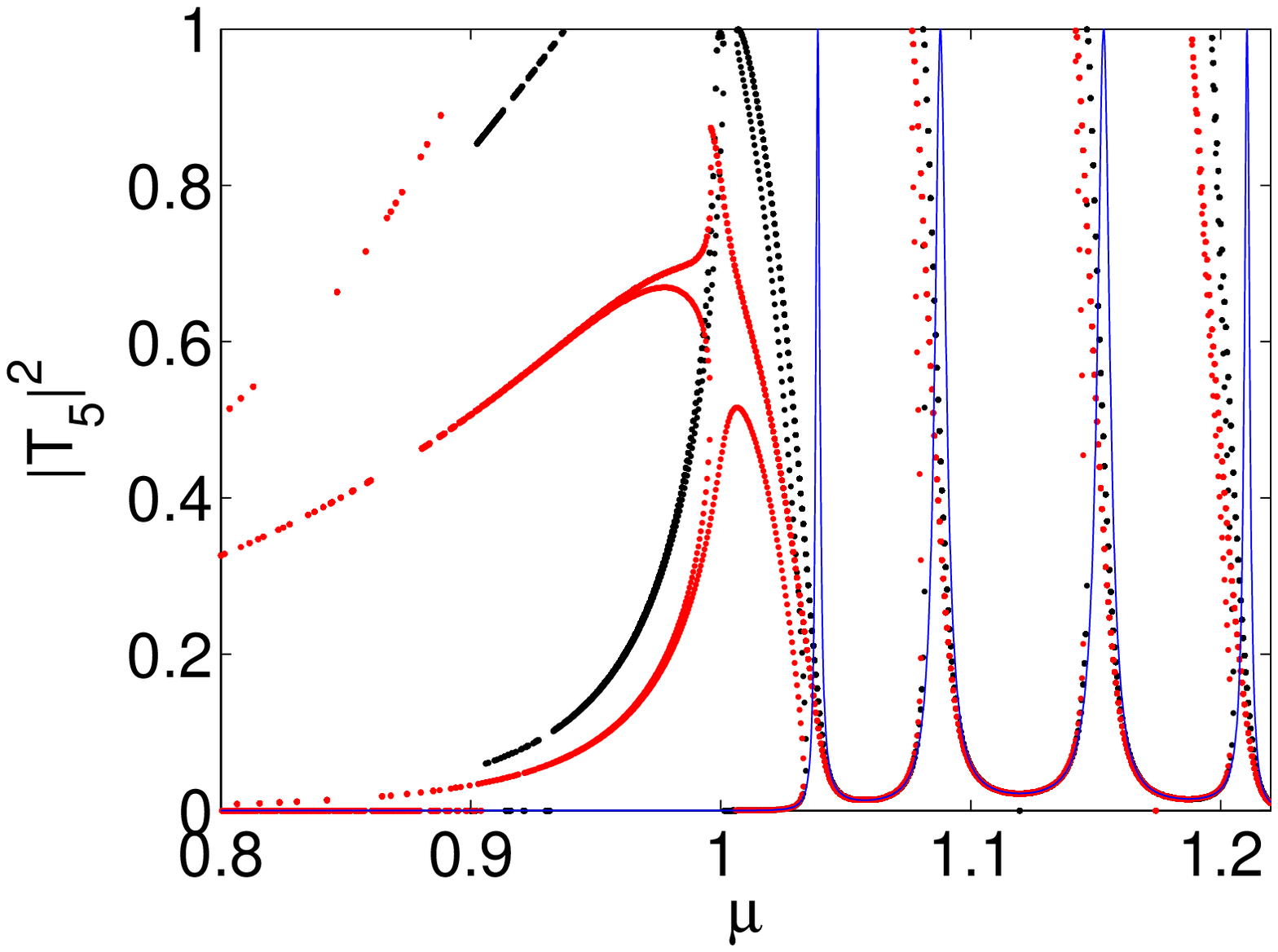}
\includegraphics[width=7.5cm,  angle=0]{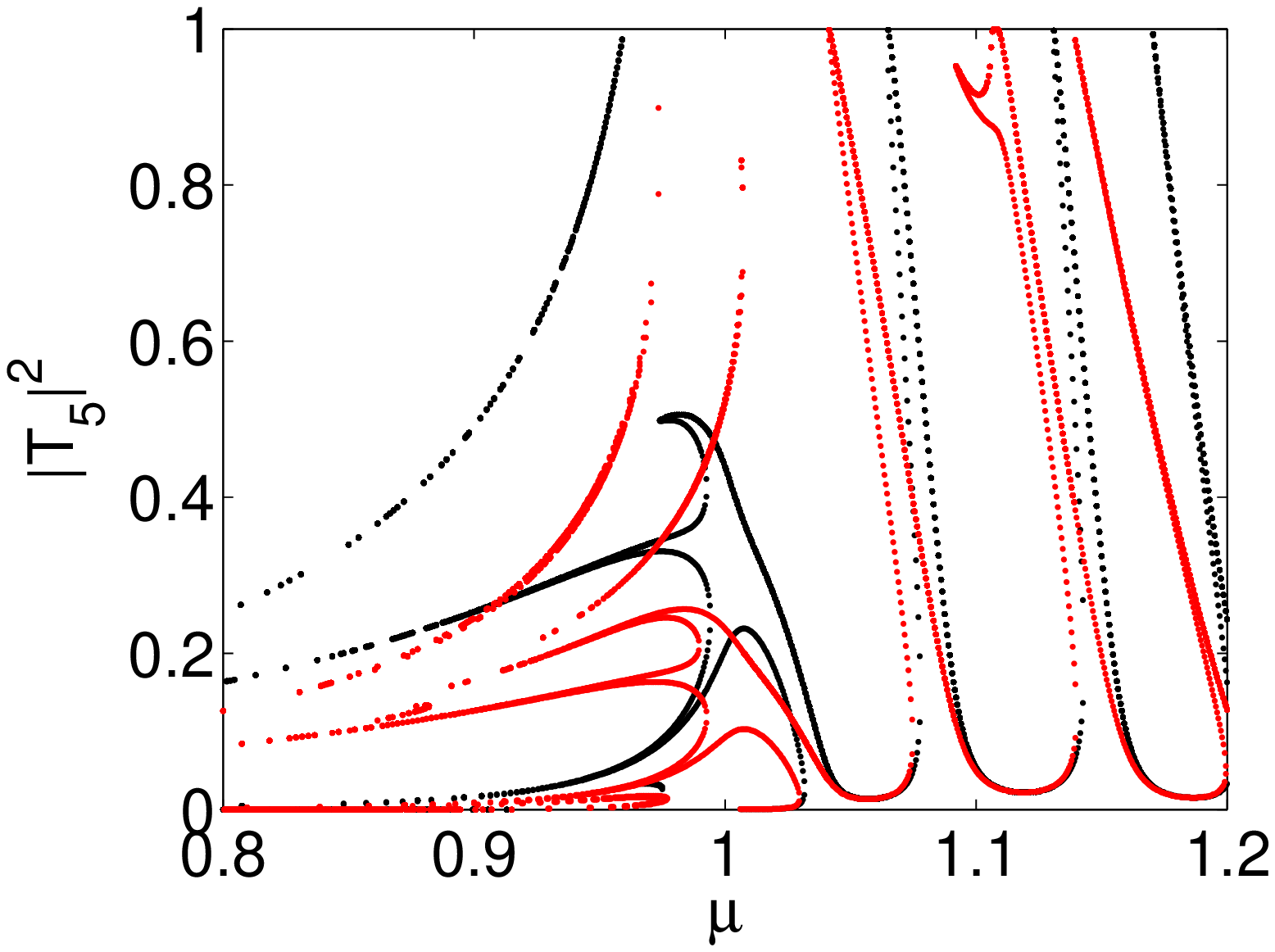}
\caption{\label{fig-posT5} {Transmission coefficient $|T_5|^2$ in dependence on the chemical potential $\mu$ for $\lambda=10$, $d=2$, $A=0.1$. Upper left panel: $g=0$ (solid blue line), $g=0.008$ (black dots), $g=0.013$ (red dots). Upper right panel: $g=0.02$ (black dots), $g=0.05$ (red dots). Lower left panel: $g=0$ (solid blue line), $g=-0.02$ (black dots), $g=-0.01$ (red dots). Lower right panel: $g=-0.04$ (black dots), $g=-0.1$ (red dots).
 }}
\end{figure}

Adding another delta function we arrive at the quintuple ($n=5$) barrier with the parameters $\lambda=10$, $d=2$ and $A=0.1$ (cf.~figure  \ref{fig-linT2_5}). The respective transmission coefficient $|T_5|^2$ in dependence on $\mu$ is shown in figure \ref{fig-posT5} for several values of the interaction constant $g$. Similar to the case of the quadruple barrier (figure \ref{fig-posT4}) increasing $g$ leads to the formation of an unconnected structure on the right hand side of the first group of resonances, its unification with the highest resonance within this group as well as to the emergence of more unconnected branches. In addition, the second resonance peak of the group develops into a fork of tree subpeaks. Again, the system reveals a completely analogous behaviour for attractive interactions $g<0$.

%-------------------------------------------------------------------------------------------------------
\section{Nonlinear oscillator model}
\label{sec_Multi_NLO}

\begin{figure}[htb]
\centering
\includegraphics[width=7.5cm,  angle=0]{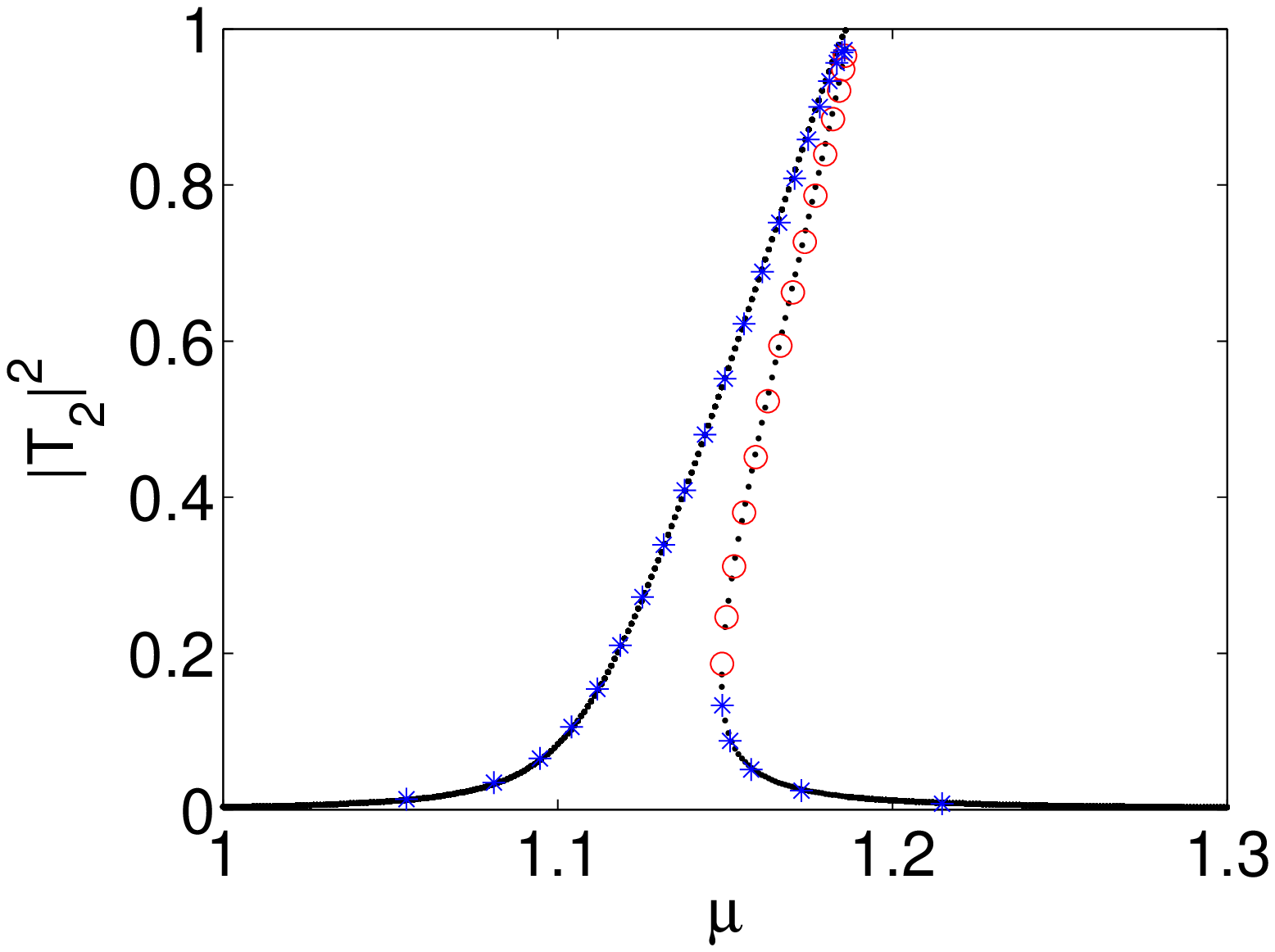}
\includegraphics[width=7.5cm,  angle=0]{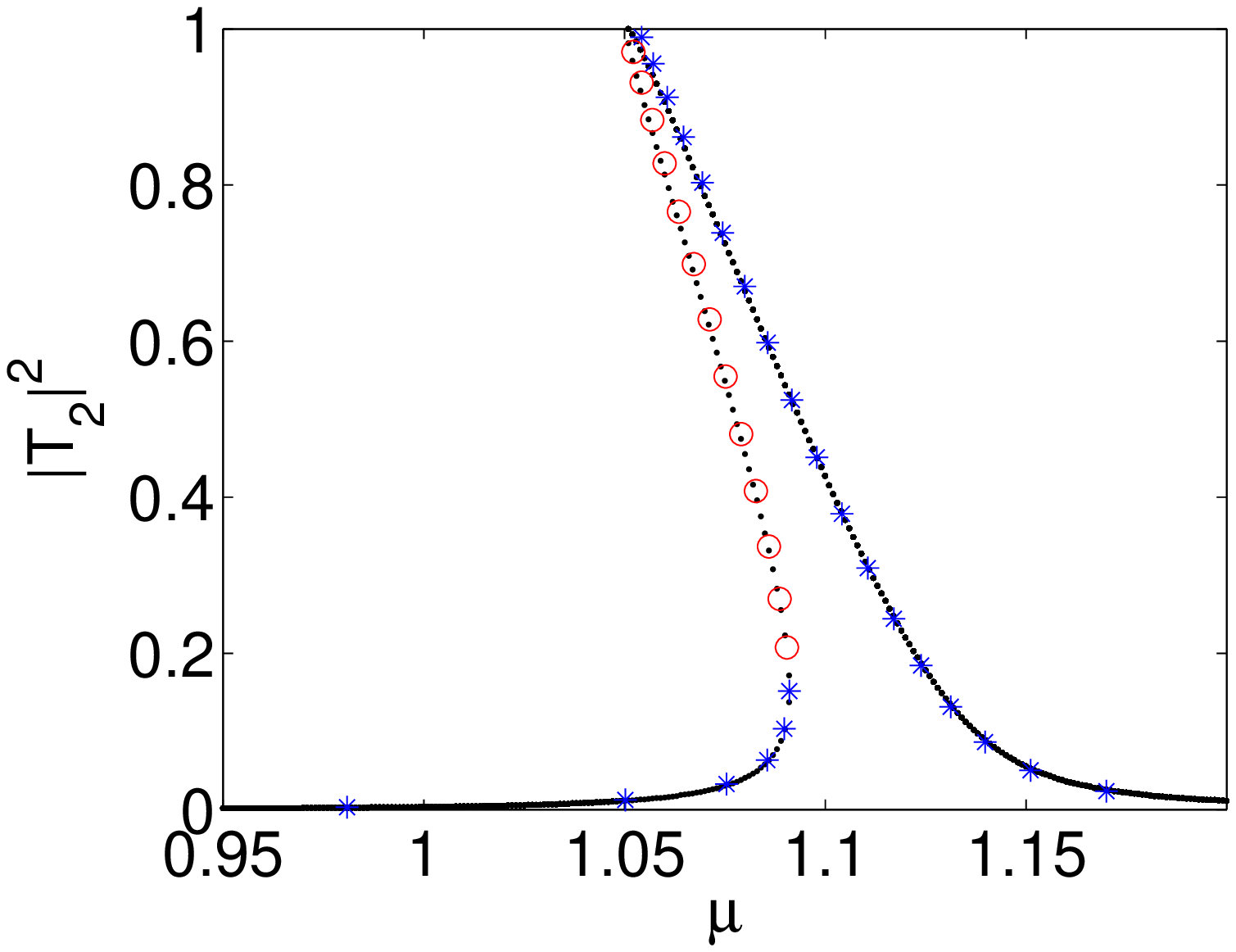}\\
\caption{\label{fig-NLO_T2} {Transmission coefficient $|T_2|^2$ in dependence on the chemical potential $\mu$ for $\lambda=10$, $d=2$, $A=0.1$. Left panel:  $g=0.05$. Right panel: $g=-0.05$. The stability predictions of the transfer map approach are indicated by black dots, the results of the nonlinear oscillator model by blue asterisks (stable regions) and red circles (unstable regions).
}}
\end{figure}
In \cite{Paul07b,08nlLorentz} it was shown that nonlinear resonant tunnelling can be understood in terms of Siegert resonances, i.e.~in the vicinity of a resonance the wavefunction $\psi(x)$ of the system is approximately given by a so-called skeleton wavefunction $\psi_{\sk}(x)$ which satifies purely outgoing (Siegert) boundary conditions and yields a complex eigenvalue $\mu_\sk-\ri \Gamma_\sk/2$. Thus we only take into account the resonant contribution to transmission neglecting the non-resonant contribution originating from sequential single barrier tunnelling. 

\begin{figure}[htb]
\centering
\includegraphics[width=7.5cm,  angle=0]{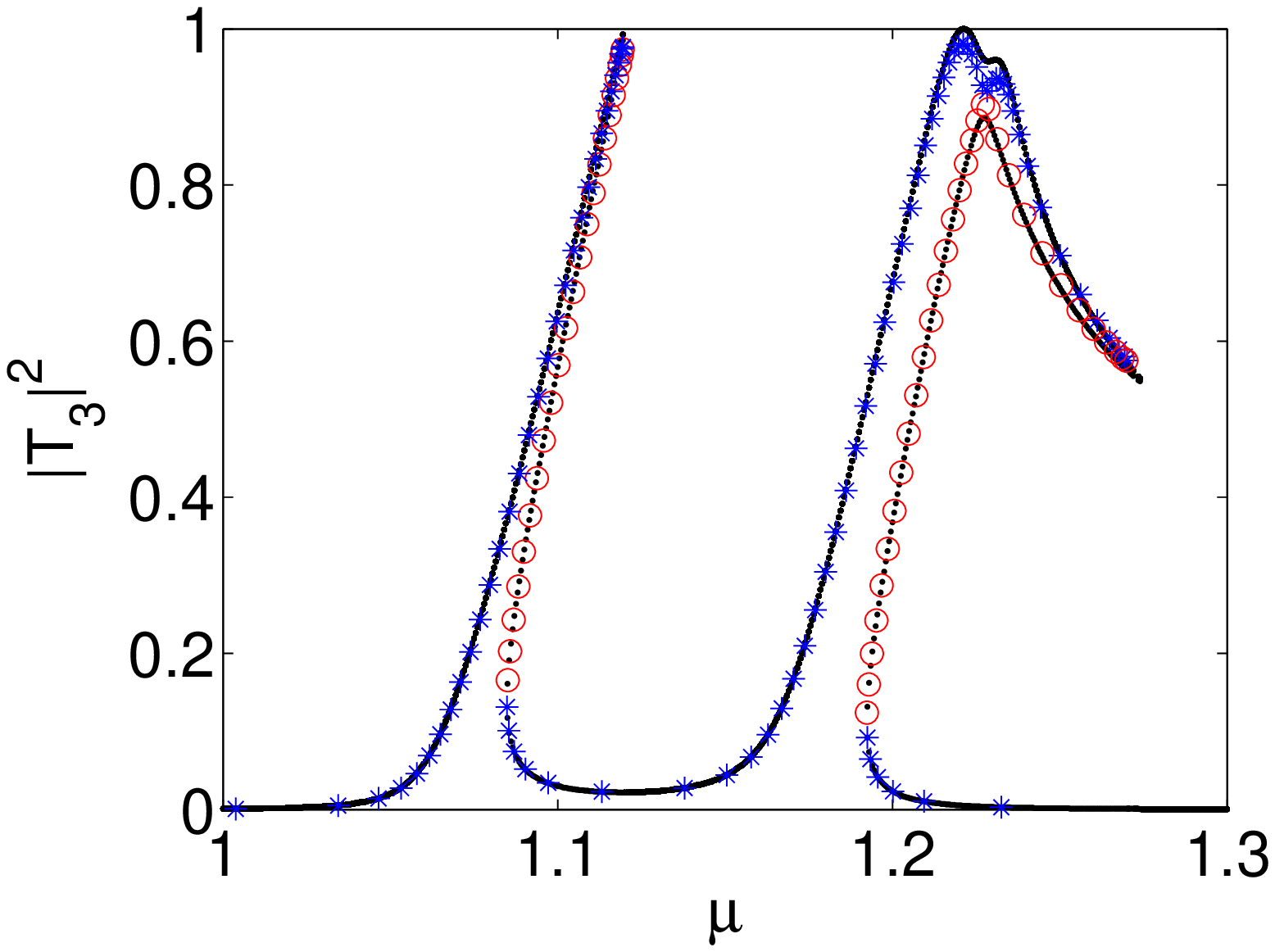}
\includegraphics[width=7.5cm,  angle=0]{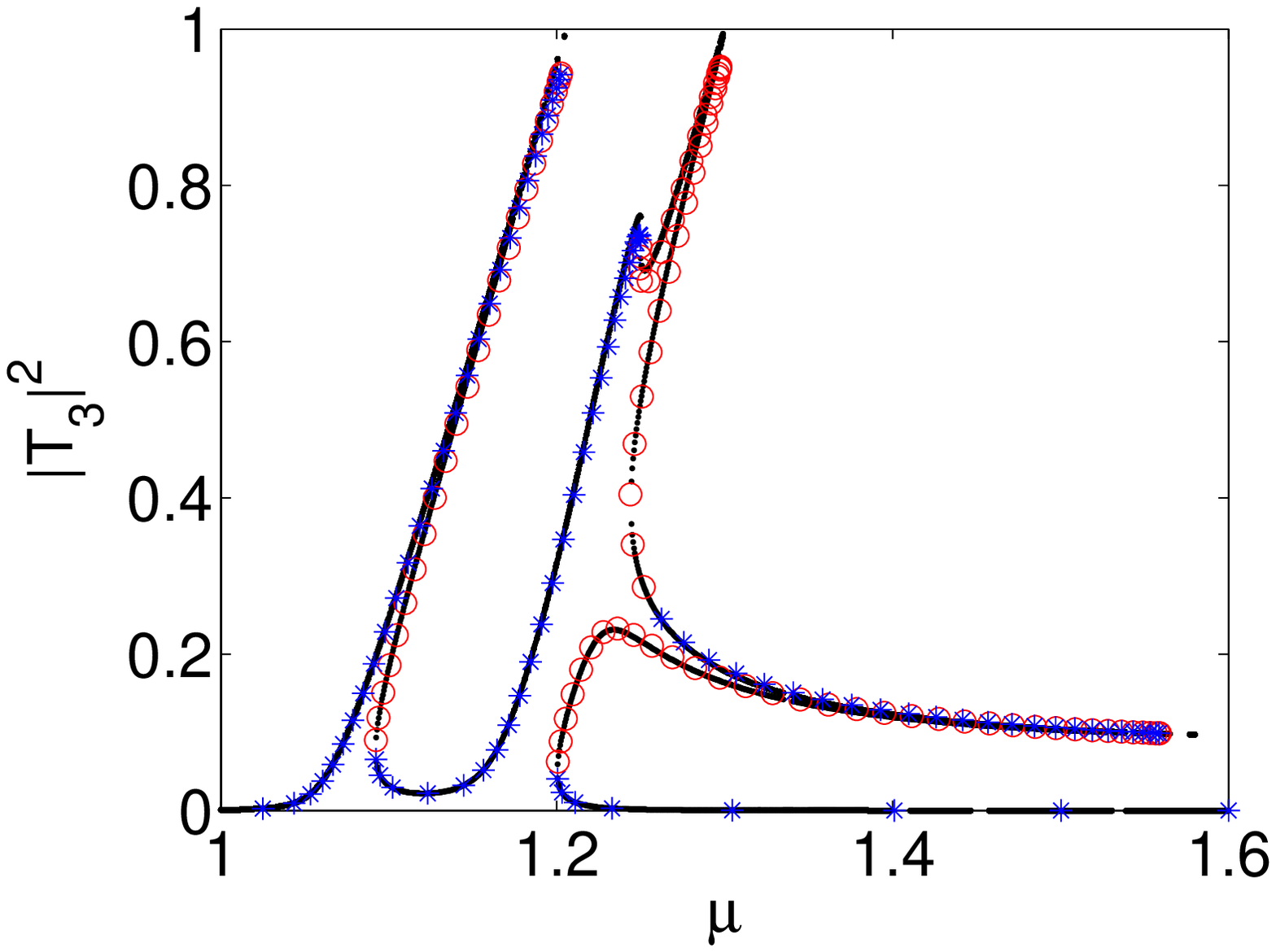}\\
\includegraphics[width=7.5cm,  angle=0]{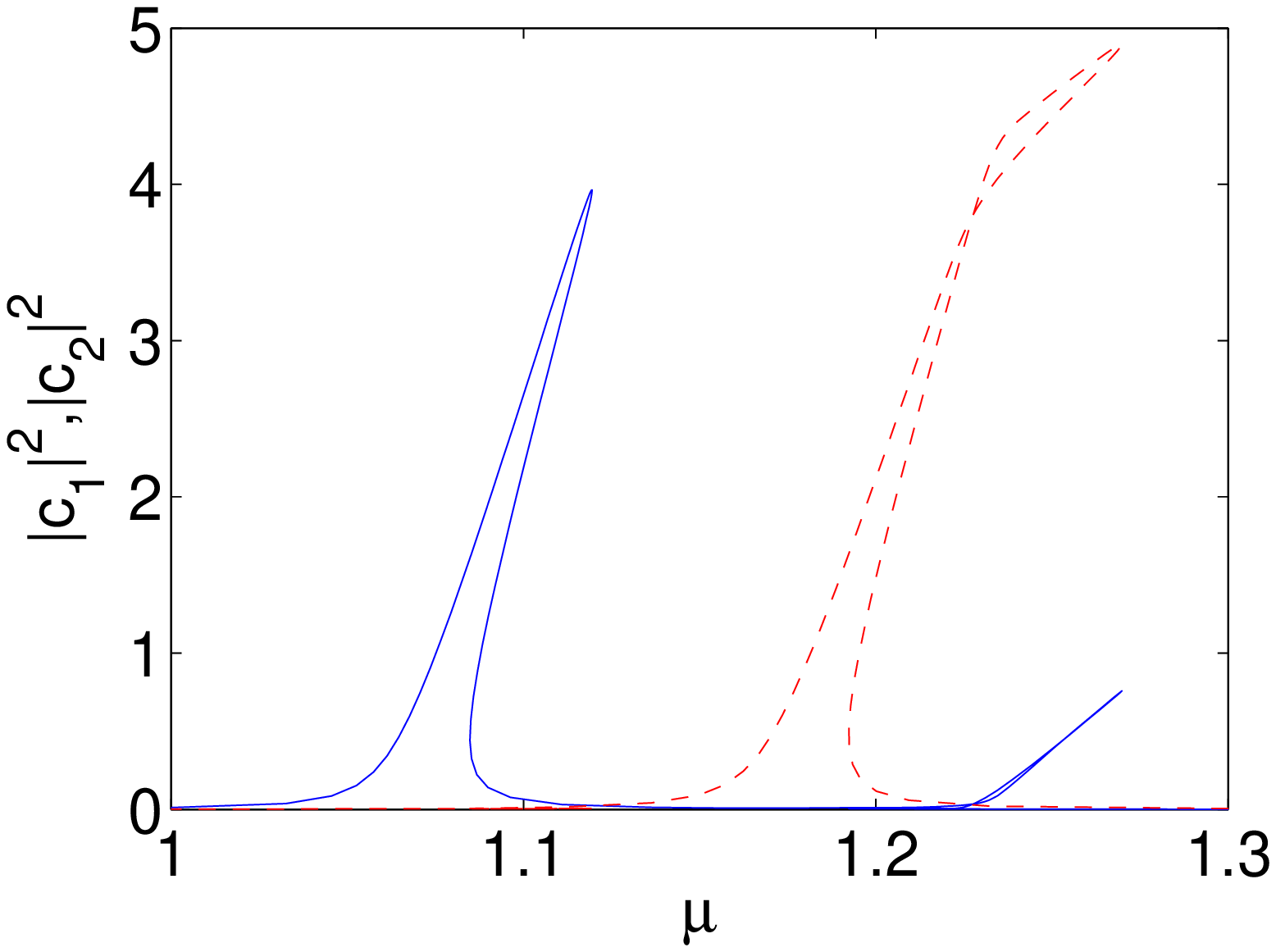}
\includegraphics[width=7.5cm,  angle=0]{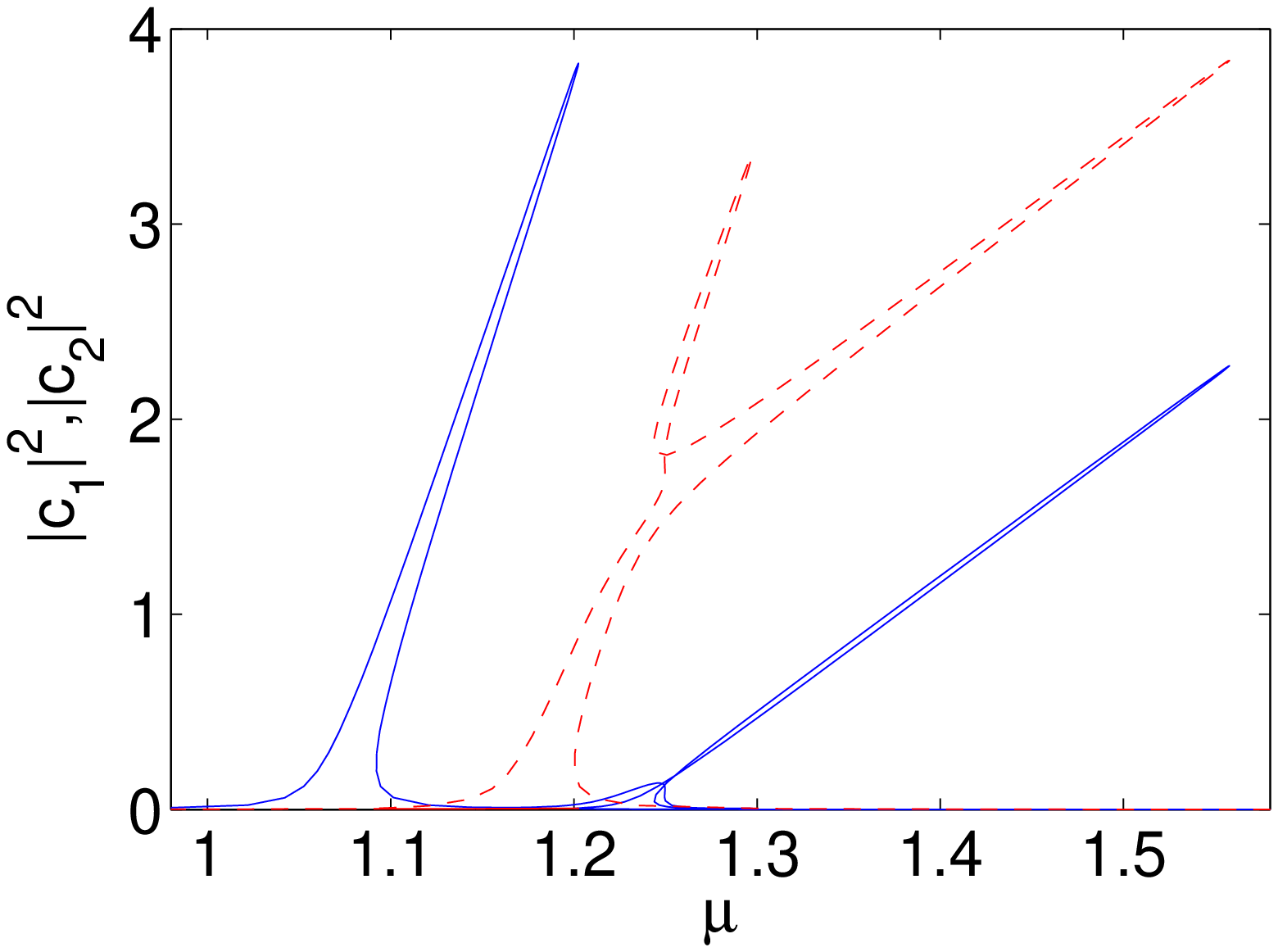}
\caption{\label{fig-NLO_T3} {Upper panels: Transmission coefficient $|T_3|^2$ in dependence on the chemical potential $\mu$ for $\lambda=10$, $d=2$, $A=0.1$. Left:  $g=0.0366$. Right: $g=0.1$. The results of the transfer map approach are indicated by black dots, the stability predictions of the nonlinear oscillator model by  blue asterisks (stable regions) and red circles (unstable regions). Lower panels: corresponding occupation numbers of the ground mode $|c_1|^2$ (solid blue line) and the first excited mode $|c_2|^2$ (dashed red line).
}}
\end{figure}
\begin{figure}[htb]
\centering
\includegraphics[width=7.5cm,  angle=0]{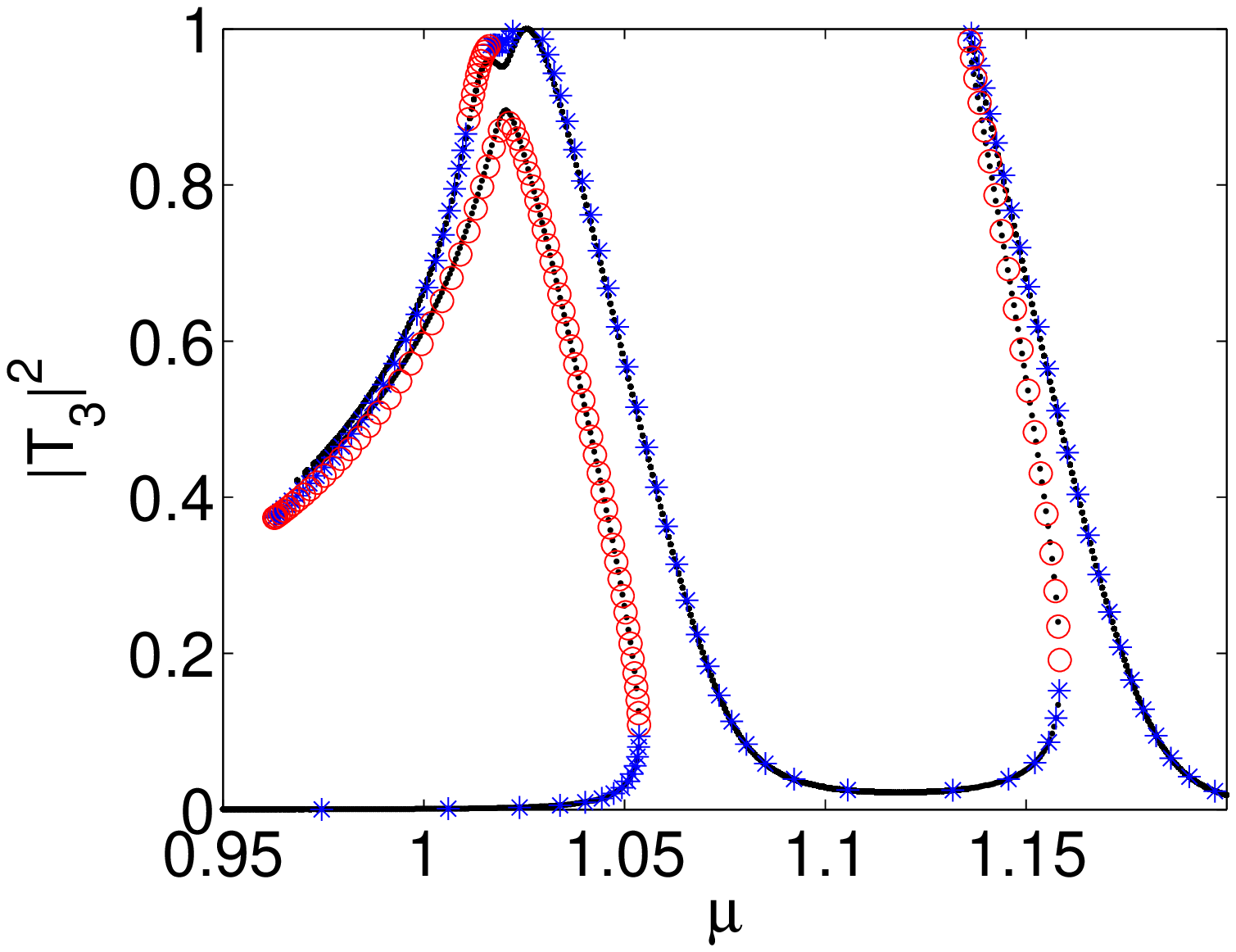}
\includegraphics[width=7.5cm,  angle=0]{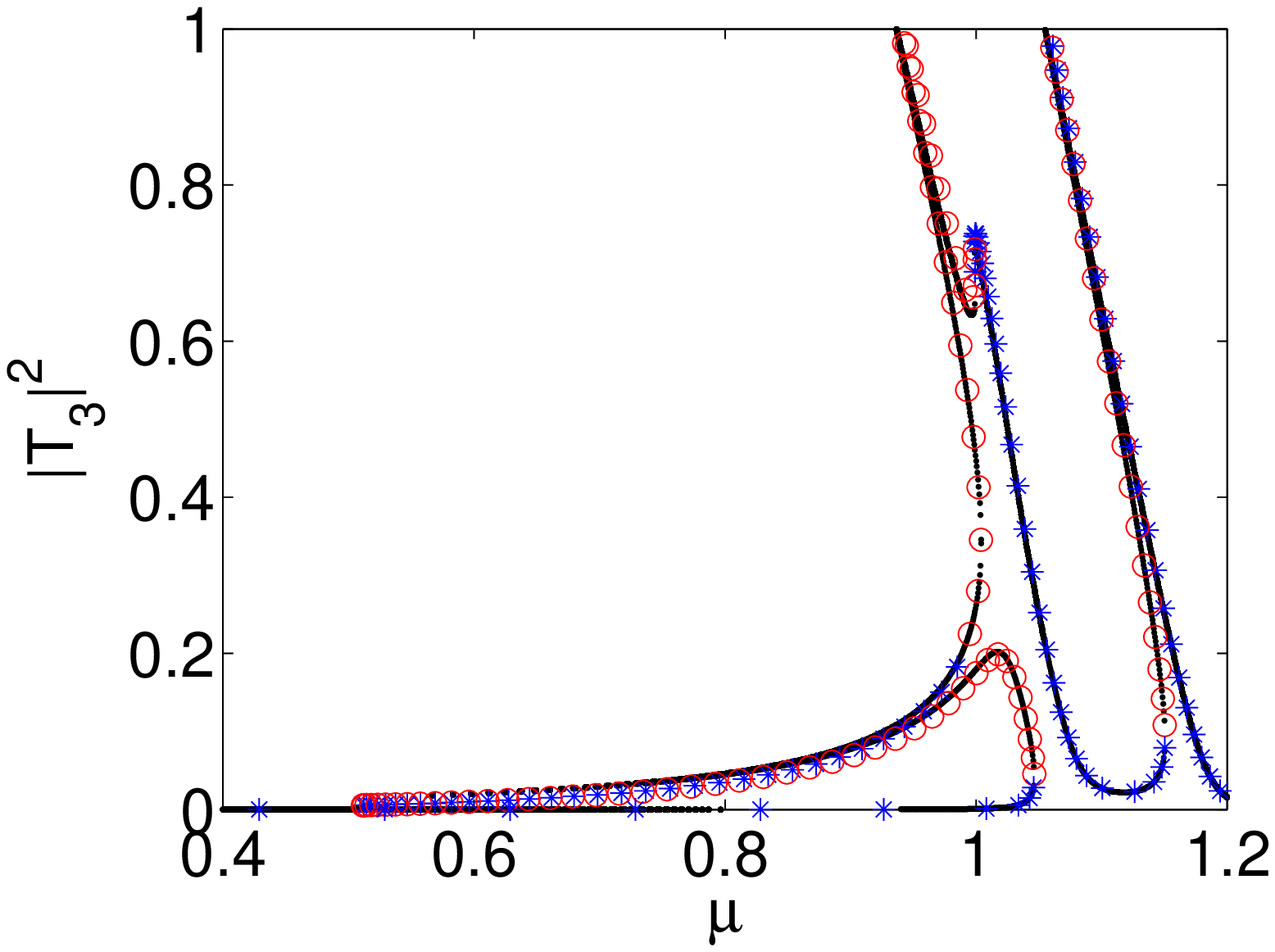}\\
\includegraphics[width=7.5cm,  angle=0]{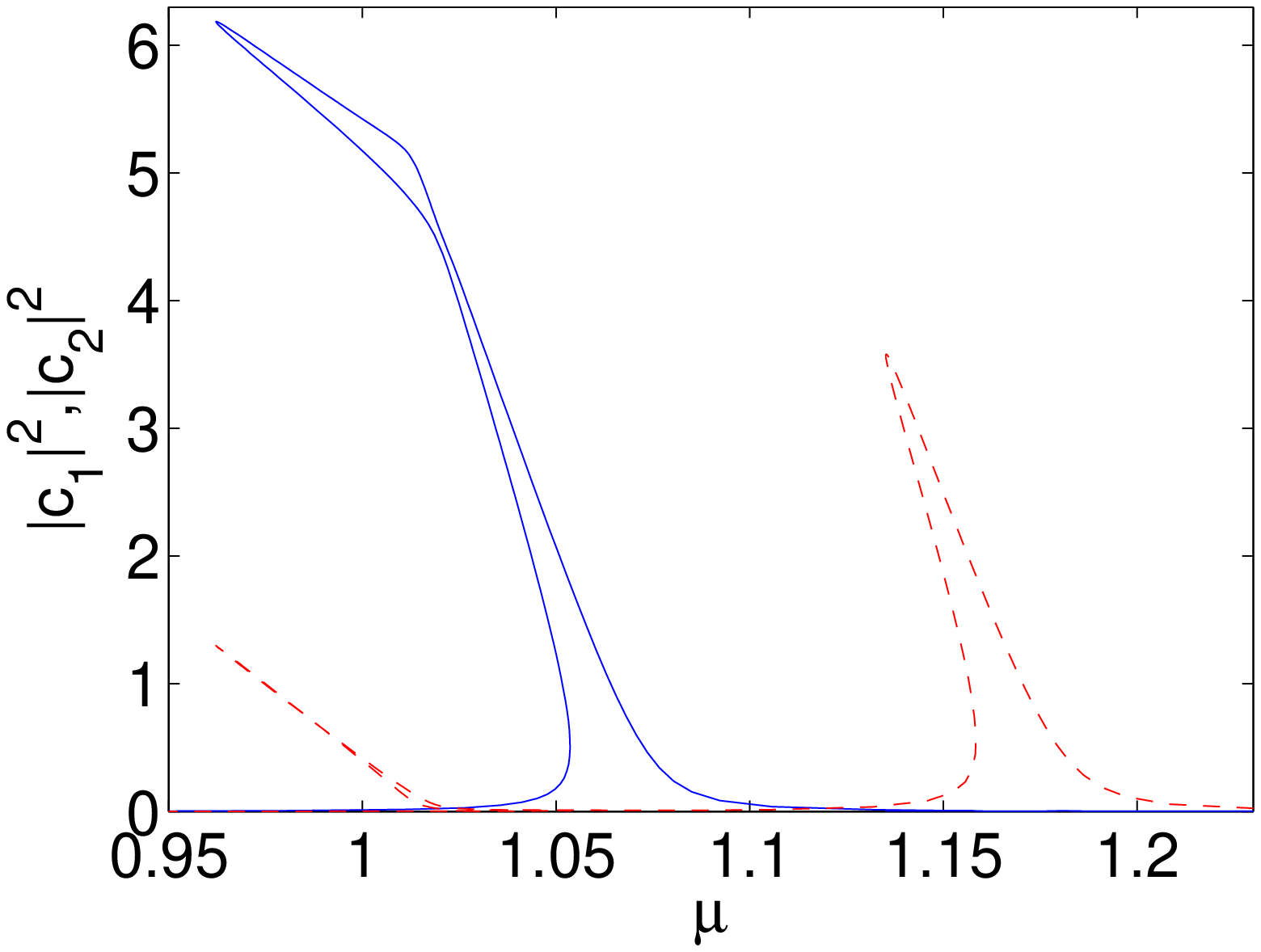}
\includegraphics[width=7.5cm,  angle=0]{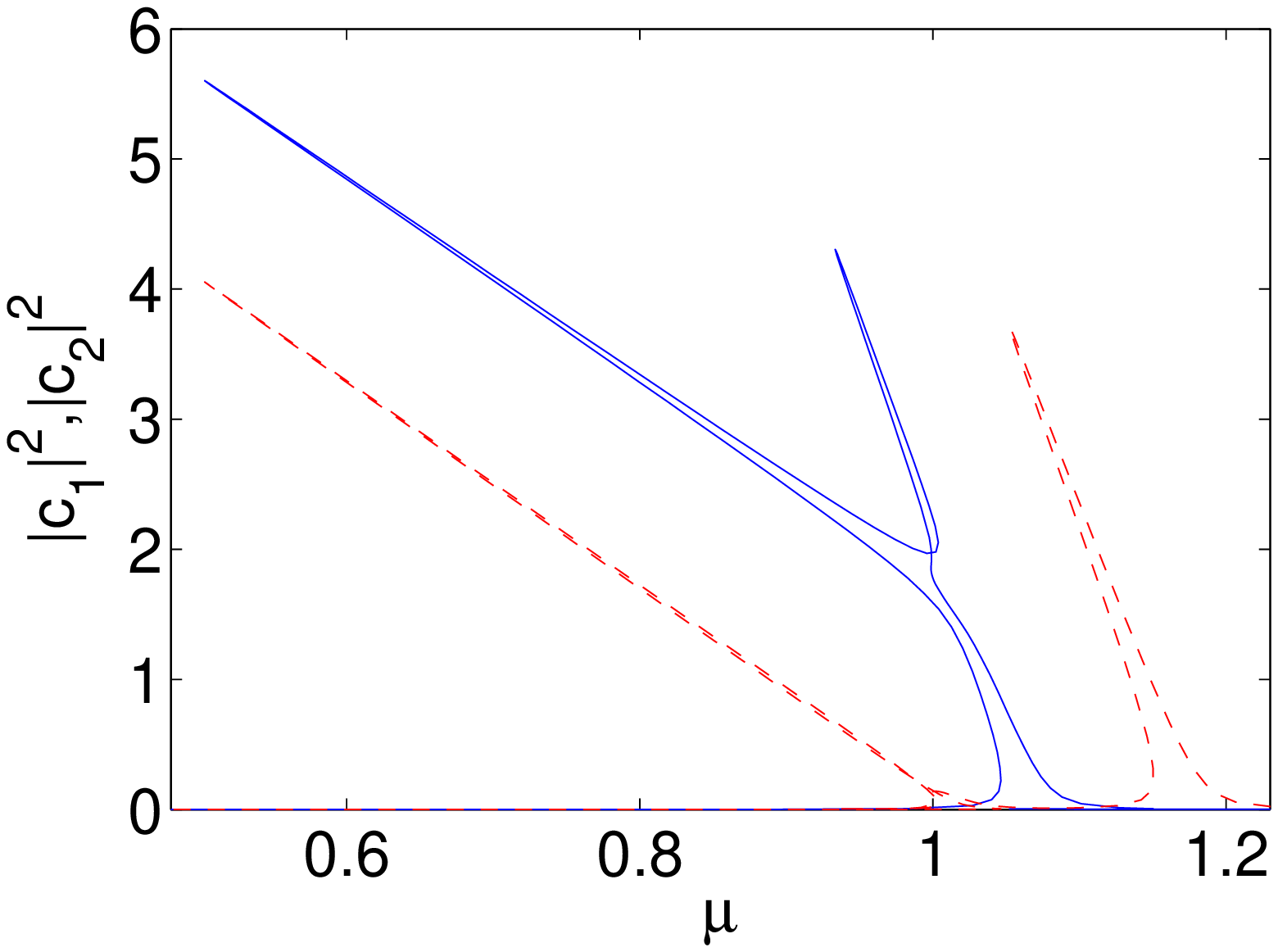}
\caption{\label{fig-NLO_Tm3} {Upper panels: Transmission coefficient $|T_3|^2$ in dependence on the chemical potential $\mu$ for $\lambda=10$, $d=2$, $A=0.1$. Left:  $g=-0.03$. Right: $g=-0.09$. The results of the transfer map approach are indicated by black dots, the stability predictions of the nonlinear oscillator model by  blue asterisks (stable regions) and red circles (unstable regions). Lower panels: corresponding occupation numbers of the ground mode $|c_1|^2$ (solid blue line) and the first excited mode $|c_2|^2$ (dashed red line).
}}
\end{figure}
This approximation can be incorporated in the time--dependent description of nonlinear resonant tunnelling mebtioned in the introduction where a source term is used to model the injection of an incoming coherent matter wave with chemical potential $\mu$. 
Inserting the ansatz $\psi(x,t)=\exp(-\ri \mu t/\hbar)\psi_\sk(x)$ into equation (\ref{GPE_t}) yields
\be
   (H_0-\mu+g|\psi_\sk(x)|^2)\psi_\sk(x)+\ri f_0 \delta(x-x_0)=0
   \label{multi_source}
\ee
where we have chosen a constant source strength $f(t)=f_0$ located at some position $x_0 \le 0$.
In analogy to \cite{09ddshell} we expand the skeleton wavefunction $\psi_\sk(x)$ in a Galerkin-type ansatz
\be
   \psi_\sk(x)=\sum_{j=1}^{n_{\rm B}}c_j u_j(x)
   \label{multi_skel}
\ee
using the first $n_{\rm B}$ eigenfunctions $\{u_j\}$ and respective eigenvalues $\{\mu_j-\ri \Gamma_j/2\}$ of the linear $(g=0)$ system
\be
    H_0=-\frac {\hbar^2}{2m}\partial_x^2+V(x) \label{H0_u}
\ee
with Siegert boundary conditions. The eigenfunctions are made square-integrable by means of exterior complex scaling (see appendix \ref{app_Multi_lin}). To calculate the stationary states we insert the ansatz (\ref{multi_skel}) into equation (\ref{multi_source}) and consider its projections
\be
 \fl  c_j(\mu_j-\ri \Gamma_j/2-\mu)+g \int_{-\infty}^{\infty} \rd x v_j^*(x) \left|\sum_{i=1}^{n_{\rm B}}c_i u_i(x)\right|^2 \sum_{l=1}^{n_{\rm B}}c_l u_l(x)+ \ri f_0 v_j^*(x_0)=0
   \label{multi_proj}
\ee
on the $n_{\rm B}$ left eigenvectors $\{v_j\}$ of $H_0$. The $n_{\rm B}$ nonlinear equations (\ref{multi_proj}), which determine the $n_{\rm B}$ coefficients $\{c_j\}$, are solved with a Newton algorithm. Obviously all equations decouple in the noninteracting case $g=0$. A system of nonlinear coupled oscillators similar to the one described by equation (\ref{multi_proj}) has been investigated in the context of micromechanical and nanomechanical resonator arrays \cite{Lifs03}.

Since the transmission coefficient for a potential with $n$ barriers shows groups of $n-1$ resonances we use $n_{\rm B}=n-1$ basis functions to compute the transmission coefficient in the vicinity of the first group of resonances.

The source strength $f_0$ is connected with the incoming wave amplitude $A$ via $f_0=\hbar^2\ri k A/m$ (cf.~\cite{Paul05,Paul07b,08nlLorentz}) with $k=\sqrt{2m \mu}/\hbar$.
For simplicity we choose $x_0=0$.
The transmission coefficient is given by the solutions of (\ref{multi_proj}) via
\be
  |T|^2=\frac{j_\rt}{j_{\rm in}}
\ee
where
\be
   j_\rt=-\frac{\ri \hbar}{2m} \left(\psi_\sk^* \psi_\sk'-\psi_\sk {\psi_\sk^*}' \right)\big|_{x=(n-1)d}
\ee
and $j_{\rm in}=\hbar k |A|^2/m$.
In this resonance ansatz the system is described by a small number of square integrable functions rather than by a continuum of distributions which can be favourable in many situations. %It might even provide an alternative way to treat
Another advantage lies in the fact that the stability of a stationary solution can be analyzed in a straightforward way (see below).

For illustration we have a closer look at the special case of a single mode, i.~e.~$n_{\rm B}=1$ (and thus $n=n_{\rm B}+1=2$), which models tunnelling through a single well/double barrier structure. Equation (\ref{multi_proj}) now reads
\be
  c_1(\mu_1-\ri \Gamma_1/2-\mu)+g w_{11}^{11}|c_1|^ 2c_1+ \ri f_0 v_1^*(x_0)=0 
  \label{Multi_c1}
\ee
with $w_{11}^{11}=\int_{-\infty}^{\infty} v_1^*(x)u_1^*(x)u_1(x)u_1(x)\, \rd x$. % (cf.~notation in chapter \ref{chap-DDShell}).
The squared magnitude of equation (\ref{Multi_c1})
\be
   |c_1|^2=\frac{|f_0|^2 \, |v_1(x_0)|^2}{(\mu1+g {\rm Re}(w_{11}^{11})|c_1|^2-\mu)^2+(\Gamma_1/2+g{\rm Im}(w_{11}^{11})|c_1|^2)^2}
   \label{Multi_c1Q}
\ee
provides a self-consistent equation for the occupation number $|c_1|^2$ of the basis function $u_1(x)$. Note that, due to symmetry, $|v_1(x_0)|^2=|u_1(x_0)|^2=|u_1(d+|x_0|)|^2$. By means of the Siegert formula $|u_1(x_0)|^2$ can be expressed in terms of the decay coefficient $\Gamma_1$ via
\be
   \Gamma_1/2=\frac{\hbar^2 k}{m}\frac{|u_1(x_0)|^2}{\int_{x_0}^{d+|x_0|}\rd x \, |u_1(x)|^2}\approx\frac{\hbar^2 k}{m}|u_1(x_0)|^2
   \label{Multi_Gamma1}
\ee
with $\int_{x_0}^{d+|x_0|}\rd x \, |u_1(x)|^2 \approx 1$ and $k=\sqrt{2m \mu}/\hbar$.
In order to express equation (\ref{Multi_c1Q}) in terms of $|T|^2=j_\rt/j_{\rm in}$ instead of $|c_1|^2$ we evaluate the current density $j_\rt$ at $x=d+|x_0|$ which yields $j_\rt=\frac{\hbar k}{m}|u_1(x_0)|^2|c_1|^2$. Using equation (\ref{Multi_Gamma1}) we obtain
$j_\rt =\hbar |c_1|^2 \Gamma_1/2$.
The transmission coefficient is thus given by
\be
  |T|^2=\frac{j_\rt}{j_{\rm in}}=\frac{m \Gamma_1}{2 \hbar^2k |A|^2}|c_1|^2 \, .
  \label{Multi_Tq_c1Q}
\ee
Using equations (\ref{Multi_Gamma1}), (\ref{Multi_Tq_c1Q}) and $|f_0|^2=\hbar^4k^2|A|^2/m$ equation (\ref{Multi_c1Q}) can be written as
\be
   |T|^2=\frac{\Gamma_1^2/4}{(\mu-\mu_\sk)^2+\Gamma_\sk^2/4}
   \label{Multi_nl_L1}
\ee
with the skeleton curves
\begin{eqnarray}
\mu_\sk(|T|^2)&=&\mu_1+g {\rm Re}(w_{11}^{11})\frac{2 |A|^2 \hbar^2 k}{m \Gamma_1}|T|^2 \label{Multi_mu_sk}\\
\Gamma_\sk(|T|^2)/2&=&\Gamma_1/2+g {\rm Im}(w_{11}^{11})\frac{2 |A|^2 \hbar^2 k}{m \Gamma_1}|T|^2 \label{Multi_Gamma_sk} \, .
\end{eqnarray}
%In the vicinity of the resonance where $\mu \approx \mu_\sk$ we make the approximations $k \approx k_\sk=\sqrt{2m \mu_\sk)}/\hbar$
%and $\Gamma_1 \approx \Gamma_\sk$ so that we finally obtain the nonlinear Lorentz profile
For small values of ${\rm Im}(w_{11}^{11}$ we can make the aproximation $\Gamma_1 \approx \Gamma_\sk$ in the numerator of equation (\ref{Multi_nl_L1}) arriving at the nonlinear Lorentz profile
\be
   |T|^2 \approx \frac{\Gamma_\sk^2/4}{(\mu-\mu_\sk)^2+\Gamma_\sk^2/4}
   \label{Multi_nl_Lsk}
\ee
derived in \cite{Paul07b,08nlLorentz}. Here the skeleton curves (\ref{Multi_mu_sk}) and (\ref{Multi_Gamma_sk}) are given in first order approximation in $|T|^2$.

In order to perform a linear stability analysis of a stationary solution $\psi_\sk$ we insert $\psi(t)=\left(\psi_\sk+\delta\psi(t)\right)\exp(-\ri \mu t/\hbar)$ into equation (\ref{GPE_t}) and retain only terms linear in $\delta \psi$. The spectral decomposition $\delta \psi(t)= \chi_- \exp(-\ri \omega t)+\chi_+^*\exp(\ri \omega^*t)$  then leads to the Bogoliubov-de Gennes equations 
\be
\hbar \omega \begin{pmatrix}\chi_- \\ \chi_+ \end{pmatrix}=
\begin{pmatrix}H_{\rm GP}+g|\psi_\sk|^2-\mu & g \psi_\sk^2 \\ -g{\psi_\sk^*}^2&-H_{\rm GP}^*-g|\psi_\sk|^2+\mu\end{pmatrix}\begin{pmatrix}\chi_- \\  \chi_+ \end{pmatrix} 
\label{Multi-BdG}
\ee
with $H_{\rm GP}=H_0+g|\psi_\sk|^2$.
%In contrast to section \ref{subsec-DD_Galerkin} we do not include a projection orthogonal to the background solution $\psi_\sk(x)$ since it is an actual stationary rather than just quasistationary solution of equation (\ref{multi_source}). 
Instability occurs if there are eigenmodes with positive imaginary part since their population grows exponentially in time. The eigenvalue equation (\ref{Multi-BdG}) is solved in the dual basis $\{u_j\}$ and $\{v_j\}$.

\begin{figure}[htb]
\centering
\includegraphics[width=7.5cm,  angle=0]{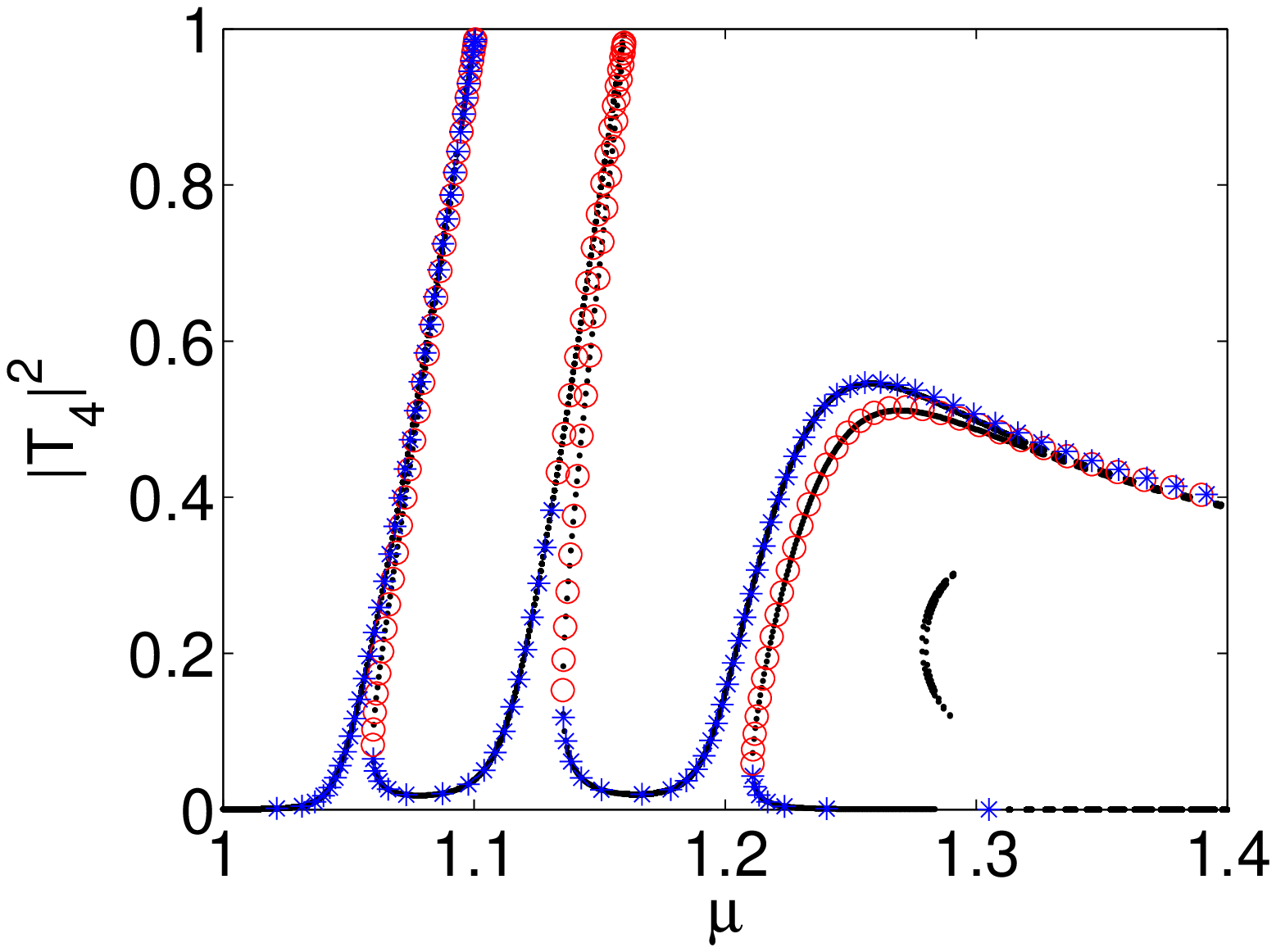}
\includegraphics[width=7.5cm,  angle=0]{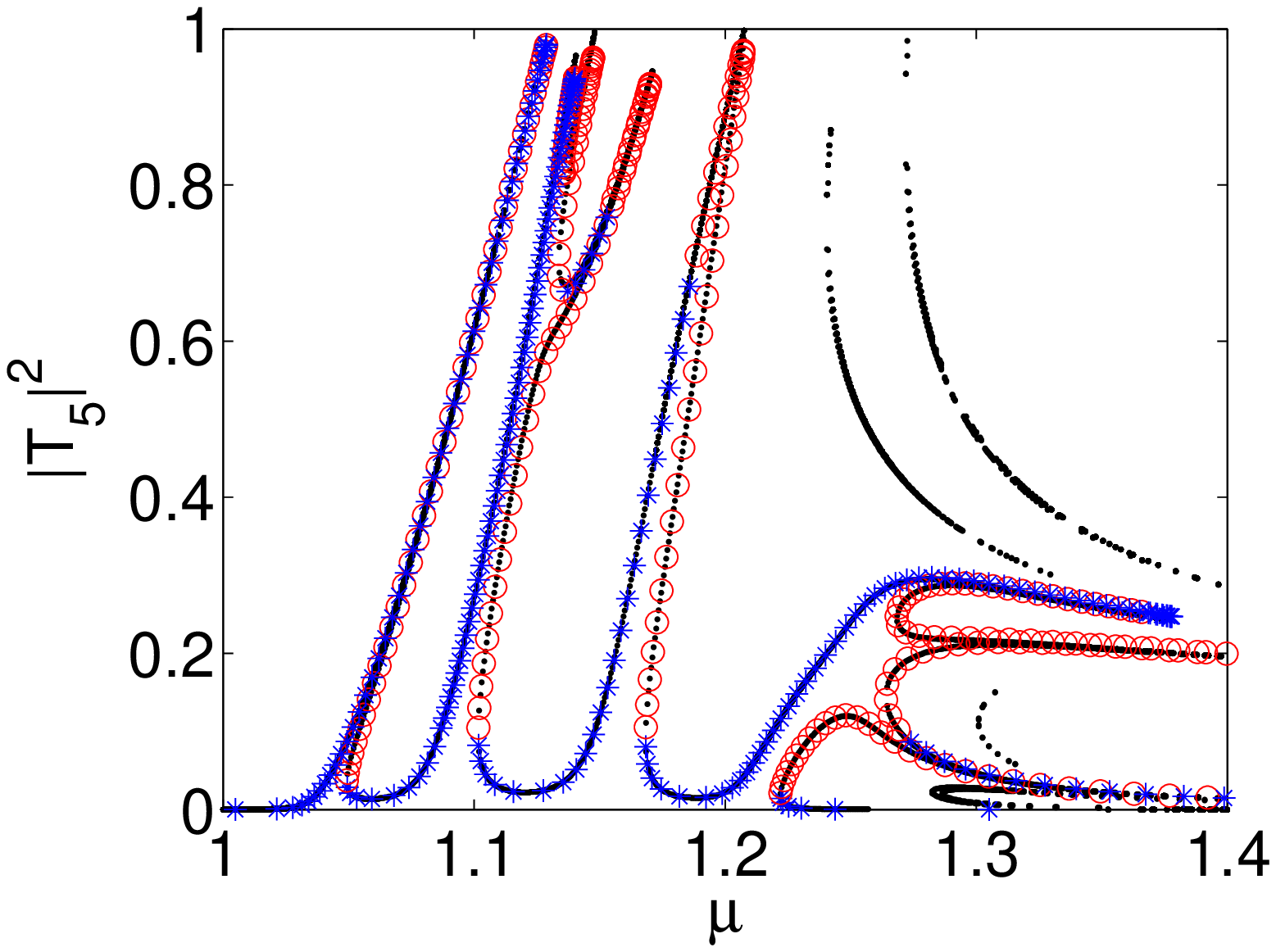}\\
\includegraphics[width=7.5cm,  angle=0]{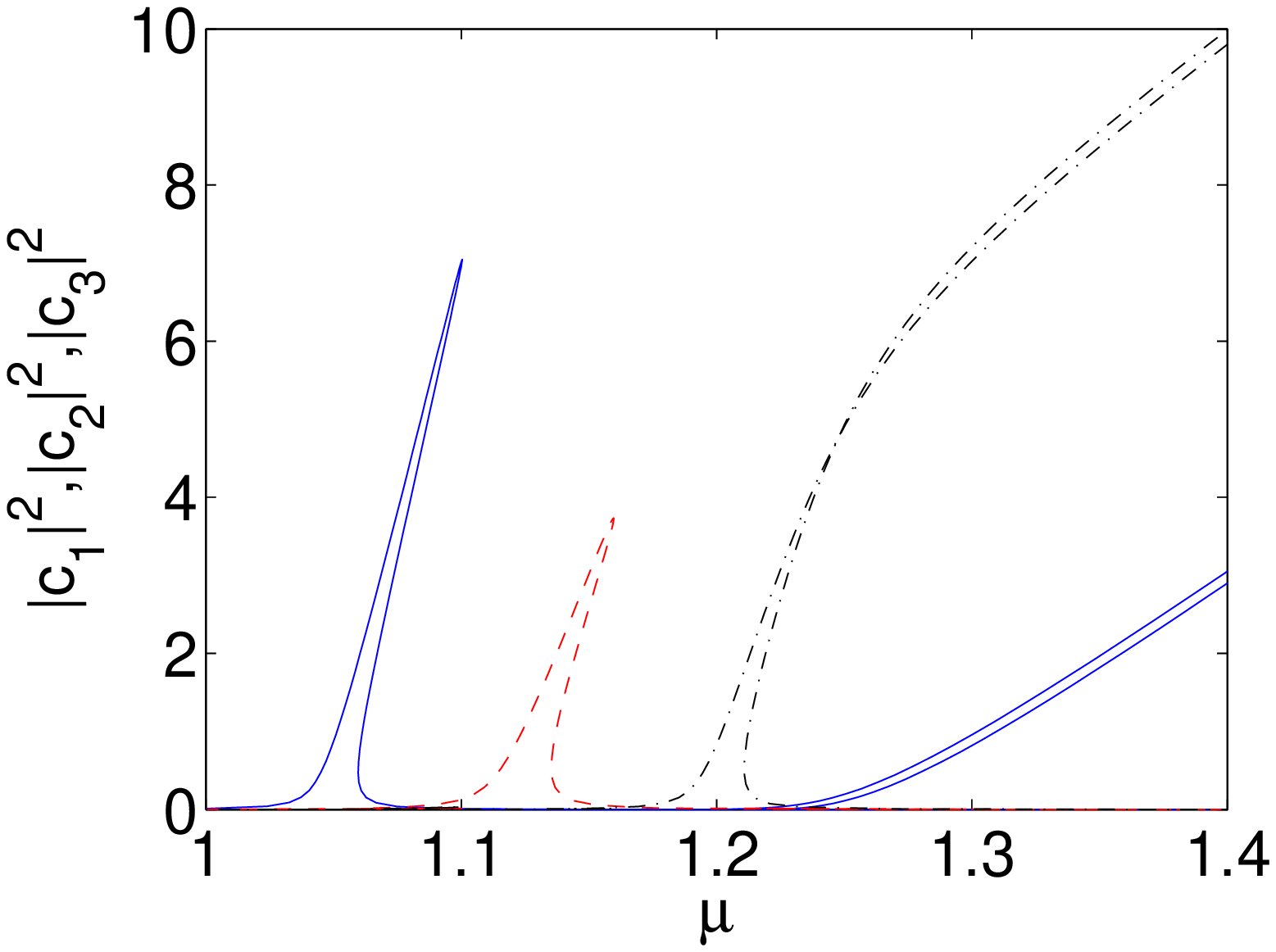}
\includegraphics[width=7.5cm,  angle=0]{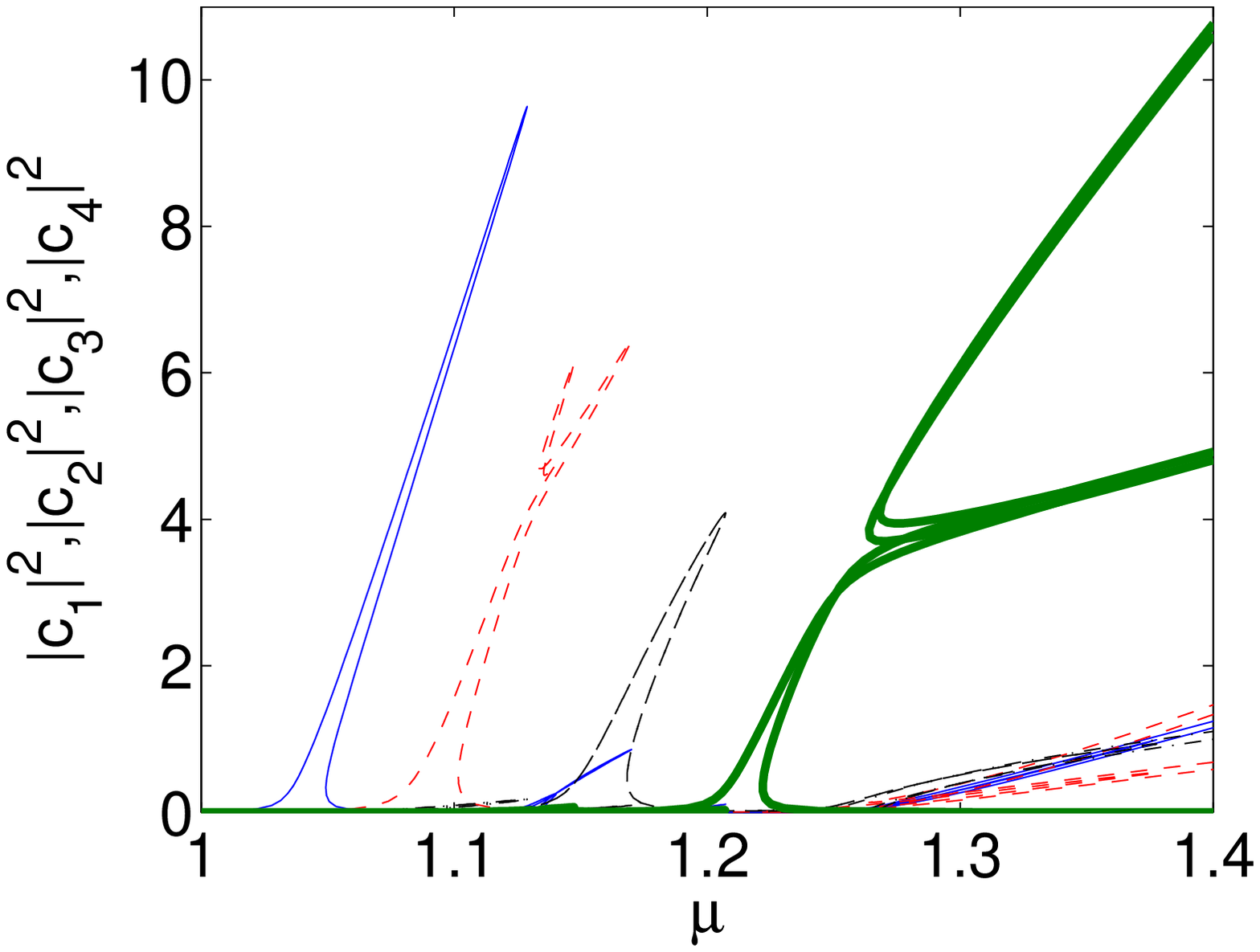}
\caption{\label{fig-NLO_T45} {Upper panels: Transmission coefficients in dependence on the chemical potential $\mu$ for $\lambda=10$, $d=2$, $A=0.1$. Left:  $|T_4|^2$ for $g=0.03$. Right: $|T_5|^2$ for $g=0.05$. The results of the transfer map approach are indicated by black dots, the stability predictions of the nonlinear oscillator model by  blue asterisks (stable regions) and red circles (unstable regions). Lower panels: corresponding occupation numbers of the ground mode $|c_1|^2$ (thin blue line), first $|c_2|^2$ (dashed red line), second $|c_3|^2$ (dashed dotted black line) and third excited mode $|c_4|^2$ (bold green line).
}}
\end{figure}

In figures \ref{fig-NLO_T2}--\ref{fig-NLO_T45} we compare the predictions of the nonlinear oscillator model with the results of the transfer map approach for different numbers of barriers and interaction constants. In all cases the agreement between both methods is quite good. Within the nonlinear oscillator approach branches which are not connected to the main part of the transmission coefficient prove difficult to find numerically and are therefore not taken into account.
The stability predictions for the double barrier (figure \ref{fig-NLO_T2}) agree with the results expected for a single parametrically driven nonlinear classical oscillator (see e.g.~\cite{Mick81}) or quantum oscillator in the classical (mean-field) limit \cite{Rigo97}.
This is also in agreement with recent numerical results for a double Gaussian barrier \cite{Erns09} where the dynamical stability of the upper branch of the transmission is explicitly demonstrated by means of a time--dependent simulation.
For more than two barriers the model predicts an increasingly complicated distribution of stable and unstable regions.
The occupation numbers $\{|c_j|^2\}$ of the modes $\{u_j\}$ shown in the lower panels of figures \ref{fig-NLO_T3}--\ref{fig-NLO_T45} indicate that the autochtonous branches of the respective transmission coefficients are mainly described by one mode only whereas the allochtonous branches are formed by superpositions of two or more modes. 
%-------------------------------------------------------------------------------------------------------
\section{Conclusion}

In this paper we considered nonlinear resonant tunnelling through sequences of $n$ identical and equally spaced delta barriers. The stationary transmission states were calculated by means of a transfer mapping approach based on the complex solutions of the free time--independent NLSE given by Jacobi elliptic functions. As observed for single well/double barrier tunnelling (see \cite{Paul05,06nl_transport}) the nonlinearity renders the transmission coefficient bistable in the vicinity of a resonance. In addition, looped structures appear, which are not connected with other branches of the transmission coefficient. If the interaction parameter $g$ is further increased these structures unite with the main part of the transmission coefficient through an inverse beak-to-beak bifurcation. A similar effect was observed in the transmission coefficient of the finite square well (see \cite{06nl_transport}) for branches of the transmission coefficient originating from bound states of the linear ($g=0$) system destabilized by interaction. Increasing the number of barriers and the nonlinearity leads to the emergence of more and more complicated structures in the transmission coefficient which result in a suppression of resonant transport. 

Comparison with a finite basis calculation based on the resonance wavefunctions of the linear system shows that the effects described above can be understood in terms of nonlinear parametrically driven coupled oscillators. The finite basis approach also offers a straightforward way to analyze the stability of different branches of the transmission coefficient by solving the corresponding Bogoliubov-de Gennes equations.
%-------------------------------------------------------------------------------------------------------

\ack
Financial support by the Deutsche Forschungsgemeinschaft (DFG) via the Graduiertenkolleg
792 "Nichtlineare Optik und Ultrakurzzeitphysik" is gratefully acknowledged.
%-------------------------------------------------------------------------------------------------------
\begin{appendix}
\section{Left and right resonance eigenfunctions in the linear limit}
\label{app_Multi_lin}
The nonlinear oscillator approach in section \ref{sec_Multi_NLO} requires the computation of the resonance eigenfunctions $u(x)$ and corresponding eigenvalues $\mu-\ri \Gamma/2$ of the Hamiltonian $H_0$ given in equation (\ref{H0_u}) with the potential $V(x)=(\hbar^2/m)\lambda \sum_{j=0}^{n-1} \delta(x-jd)$ given in equation (\ref{delta_series}) which are obtained by solving the stationary Schr\"odinger equation
\be
  \left(-\frac{\hbar^2}{2m} \partial_x^2 +V(x)\, \right)u(x)=(\mu-\ri \Gamma/2)u(x)
  \label{app_SE}
\ee
with Siegert boundary conditions.
We make the ansatz
\be
u(x)= \left\{
                    \begin{array}{cl}
                      \exp(-\ri k x)   & x<0 \\
                     I_{j} \sin(k jd +\vartheta_j)   & (j-1)d \le  x < jd, \, 0<j<n-1 \\
                      \exp(\ri k x)   & x \ge (n-1)d
                    \end{array}
              \right.
\ee
with $k=\sqrt{2m(\mu-\ri\Gamma/2)}/\hbar$ which satisfies the Siegert boundary conditions \\$\lim_{x \rightarrow \pm \infty} u'(x)=\pm \ri k \, u(x)$.
The matching conditions at $x=0$ and $x=(n-1)d$ read
\be
   1=I_1 \sin(\vartheta_1) \,, \quad -\ri k = k I_1 \cos(\vartheta_1)-2\lambda \label{app_MC1}
\ee
and
\be
     k I_{n-1} \cos(\vartheta_{n-1})= \ri k-2\lambda \, . \label{app_MC2}
\ee
At $x=jd$, $ 0<j<n-1$ we obtain
\begin{eqnarray}
  I_j\sin(kjd+\vartheta_j)=I_{j+1}\sin(kjd+\vartheta_{j+1})\,, \label{app_MC3}\\
  k I_j\cos(kjd+\vartheta_j)=k I_{j+1}\cos(kjd+\vartheta_{j+1})-2 \lambda I_j\sin(kjd+\vartheta_j) \, .\label{app_MC4}
\end{eqnarray}
These equations are solved numerically for the complex quantities $k$, $\vartheta_j$ and $I_j$, $0<j <n$.

The wave function $u(x)$ diverges for $x\rightarrow \infty$ since ${\rm Im}(k)<0$. Therefore we use exterior complex scaling (see e.~g.~\cite{Mois98,Zavi04}) to make the wave function square integrable.
The $x$ coordinate is rotated by an angle $\theta_c$ from the point where the potential $V(x)$ becomes zero. In our case this reads
\be
x \rightarrow \left\{
                    \begin{array}{cl}
                      x \exp(\ri \theta_c)  & x<0 \\
                      x   & x \le 0 \le (n-1)d \\
         (n-1)d +(x-(n-1)d)\exp(\ri \theta_c)          &     x>(n-1)d
                    \end{array}
              \right. \, .
\ee
In the scaled region the Schr\"odinger equation becomes $\exp(2\ri \theta_c)u''(x)+k^2 u(x)=0$. The matching conditions (\ref{app_MC1}) and (\ref{app_MC2}) remain unaltered. For a sufficiently large rotation angle $\theta_c$ the wavefunction $u(x)$ becomes square integrable in $0 \le x < \infty$.

Since $H_0$ is symmetric, the corresponding left eigenfunctions $v(x)$ are given by $v(x)=(u(x))^*$.
We normalize the eigenstates such that $\int_0^\infty \rd x \, v^*(x) u(x) =1$.

\end{appendix}
%-------------------------------------------------------------------------------------------------------
\section*{References}
% 
%  \bibliographystyle{/home/agkorsch/tex/bibtex/bst/unsrtot}
%  \bibliography{/home/agkorsch/tex/bibtex/bib/abbrev,/home/agkorsch/tex/bibtex/bib/publko,/home/agkorsch/tex/bibtex/bib/paper60,/home/agkorsch/tex/bibtex/bib/paper70,/home/agkorsch/tex/bibtex/bib/paper80,/home/agkorsch/tex/bibtex/bib/paper90,/home/agkorsch/tex/bibtex/bib/paper00,/home/agkorsch/tex/bibtex/bib/rest}

\end{document}